\documentstyle[11pt]{article}
\begin{document}
\newcommand{\cosb}{{\em COS~B}\/}
\newcommand{\sas}{{\em SAS~2}\/}
\newcommand{\cgro}{{\em CGRO}\/}
\newcommand{\egret}{{\em EGRET}\/}
\newcommand{\glast}{{\em GLAST}\/}
\newcommand{\batse}{{\em BATSE}\/}
\newcommand{\comptel}{{\em COMPTEL}\/}
\newcommand{\osse}{{\em OSSE}\/}
\newcommand{\rosat}{{\em ROSAT}\/}

\newcommand{\sage}{{\tt SAGE}\/}
\newcommand{\like}{{\tt LIKE}\/}
\newcommand{\timevar}{{\tt timevar}\/}
\newcommand{\tjrecon}{{\tt tjrecon}\/}
\newcommand{\gismo}{{\tt gismo}\/}
\newcommand{\glastsim}{{\tt glastsim}\/}
\newcommand{\pulsar}{{\tt PULSAR}\/}
\newcommand{\spectral}{{\tt SPECTRAL}\/}

\newcommand{\gammaray}{$\gamma$-ray}
\newcommand{\gammarays}{$\gamma$-rays}
\newcommand{\Gammaray}{$\gamma$-Ray}
\newcommand{\Gammarays}{$\gamma$-Rays}
\newcommand{\perareasec}{~cm$^{-2}$~s$^{-1}$}
\newcommand{\perareasecmev}{~cm$^{-2}$~s$^{-1}$~MeV$^{-1}$}
\newcommand{\perareasecsr}{~cm$^{-2}$~s$^{-1}$~sr$^{-1}$}
\newcommand{\eg}{e.g.}
\newcommand{\ie}{i.e.}
\newcommand{\epm}{$e^\pm$}
\newcommand{\etal}{et~al.}
\newcommand{\sub}[1]{_{\rm #1}}
\newcommand{\expct}[1]{\left\langle #1 \right\rangle}
\newcommand{\radlen}{$X_0$}
\newcommand{\BO}{B$\emptyset$}
\newcommand{\pos}{$e^+$}
\newcommand{\el}{$e^-$}
\newcommand{\beamtest}{beam test}
\newcommand{\Beamtest}{Beam test}
\newcommand{\BeamTest}{Beam Test}

\def\accang{\hbox{$\theta_{67}$}}
\def\chisq{\hbox{$\chi^2$}}
\def\expsr{\hbox{$\cal E$}}
\def\htest{\hbox{$H$-test}}
\def\ipred{\hbox{$\dot E/D^2$}}
\def\ij{\hbox{$_{ij}$}}
\def\kl{\hbox{$_{kl}$}}
\def\psf{\hbox{\rm PSF}}
\def\TS{\hbox{\rm TS}}
\def\sn{\hbox{\rm S/N}}
\def\sqrtTS{\hbox{${\rm TS}^{1/2}$}}
\def\zsqr{\hbox{$Z_m^2$}}

\newcommand{\chapt}[1]{Chapter~\ref{#1}}
\newcommand{\eq}[1]{equation~(\ref{#1})}
\newcommand{\fig}[1]{Figure~\ref{#1}}
\newcommand{\sect}[1]{\S\ref{#1}}
\newcommand{\tbl}[1]{Table~\ref{#1}}
\newcommand{\app}[1]{Appendix~\ref{#1}}

\def\aj{AJ}         
\def\araa{ARA\&A}       
\def\apj{ApJ}           
\def\apjl{ApJ}      
\def\apjs{ApJS}     
\def\ao{Appl.Optics}        
\def\apss{Ap\&SS}       
\def\aap{A\&A}      
\def\aapr{A\&A~Rev.}        
\def\aaps{A\&AS}        
\def\azh{AZh}           
\def\baas{BAAS}     
\def\jrasc{JRASC}       
\def\memras{MmRAS}      
\def\mnras{MNRAS}       
\def\nature{Nature}     
\def\physrev{Phys.Rev.}     
\def\pra{Phys.Rev.A}        
\def\prb{Phys.Rev.B}        
\def\prc{Phys.Rev.C}        
\def\prd{Phys.Rev.D}        
\def\prl{Phys.Rev.Lett} 
\def\pasp{PASP}     
\def\pasj{PASJ}     
\def\qjras{QJRAS}       
\def\science{Science}       
\def\skytel{S\&T}       
\def\solphys{Solar~Phys.}   
\def\sovast{Soviet~Ast.}    
\def\ssr{Space~Sci.Rev.}
\def\thesis{Ph.D. Thesis}   
\def\zap{ZAp}           
\def\GROworkshop{in Proc. of the Gamma-Ray Observatory Science Workshop,
ed. W. N. Johnson (Greenbelt, MD: NASA)}
\def\EGRETsymp{in The Energetic Gamma-Ray Experiment Telescope (EGRET)
Science Symposium, ed. C. Fichtel \etal\ (Greenbelt, MD: NASA)}

\def\gt{\hbox{$>$}}
\def\lt{\hbox{$<$}}
\newcommand{\cross}{{\boldmath \times}}
\newcommand{\vdot}{{\boldmath \cdot}}

\def\deg{\hbox{$^\circ$}}
\def\hr{\hbox{$^{\rm h}$}}
\def\mn{\hbox{$^{\rm m}$}}
\def\sun{\hbox{$\odot$}}
\def\earth{\hbox{$\oplus$}}
\def\sq{\hbox{\rlap{$\sqcap$}$\sqcup$}}
\def\arcmin{\hbox{$^\prime$}}
\def\arcsec{\hbox{$^{\prime\prime}$}}
\def\fd{\hbox{$.\!\!^{\rm d}$}}
\def\fh{\hbox{$.\!\!^{\rm h}$}}
\def\fm{\hbox{$.\!\!^{\rm m}$}}
\def\fs{\hbox{$.\!\!^{\rm s}$}}
\def\fdg{\hbox{$.\!\!^\circ$}}
\def\farcm{\hbox{$.\mkern-4mu^\prime$}}
\def\farcs{\hbox{$.\!\!^{\prime\prime}$}}
\def\fp{\hbox{$.\!\!^{\scriptscriptstyle\rm p}$}}
\def\us{\hbox{$\mu$s}}

\def\onehalf{1/2}
\def\onethird{1/3}
\def\twothirds{2/3}
\def\onequarter{1/4}
\def\threequarters{3/4}
\def\oneeighth{1/8}
\def\ubvr{\hbox{$U\!BV\!R$}}        
\def\ub{\hbox{$U\!-\!B$}}       
\def\bv{\hbox{$B\!-\!V$}}       
\def\vr{\hbox{$V\!-\!R$}}       
\def\ur{\hbox{$U\!-\!R$}}       

\input epsf 
\author{W.B.Atwood$^{f,h}$, S.Ritz$^c$, P.Anthony$^f$, E.D.Bloom$^f$, P.E.Bosted$^a$, 
J.Bourotte$^j$, C.Chaput$^f$, \\ X.Chen$^b$, D.L.Chenette$^d$, D.Engovatov$^f$, R.Erickson$^f$, 
T.Fieguth$^f$, P.Fleury$^j$, R.Gearhart$^f$, \\ G.Godfrey$^f$, J.E.Grove$^e$, J.A.Hernando$^h$,
M.Hirayama$^h$, S.Jaggar$^h$, 
R.P.Johnson$^h$,\\ W.N.Johnson$^e$, B.B.Jones$^g$, W.Kr\"{o}ger$^h$, Y.C.Lin$^g$, 
C.Meetre$^c$, P.F.Michelson$^g$,\\ P.A.Milne$^l$, A.Moiseev$^{c,k}$, P.Nolan$^g$, 
J.P.Norris$^c$, M.Oreglia$^i$, J.F.Ormes$^{c}$,\\ B.F.Phlips$^{e,k}$, A.Pocar$^h$, 
H.F.-W.Sadrozinski$^h$,\\
Z.Szalata$^a$, D.J.Thompson$^c$, W.F.Tompkins$^g$ \\ 
\\ \\ \\ 
$^a$ American University, Washington, D.C. \\
$^b$ Columbia University, New York, N.Y. \\
$^c$ NASA Goddard Space Flight Center, Greenbelt, MD. \\
$^d$ Lockheed-Martin, Palo Alto, CA. \\
$^e$ Naval Research Laboratory, Washington, D.C.  \\
$^f$ Stanford Linear Accelerator Center, Stanford, CA. \\
$^g$ Stanford University, Stanford, CA. \\
$^h$ University of California Santa Cruz, Santa Cruz, CA. \\
$^i$ University of Chicago, Chicago, IL.\\
$^j$ Ecole Polytechnique, Palaiseau, France \\ 
$^k$ Universities Space Research Association, Washington, D.C. \\
$^l$ Clemson University, Clemson, SC.}
\title{Beam Test of Gamma-ray Large Area Space Telescope Components\\ \vskip 1cm}
\date{April 6, 1999}
\maketitle

\abstract{A beam test of GLAST (Gamma-ray Large Area Space 
Telescope) components was performed at the Stanford Linear Accelerator 
Center in October, 1997.  These beam test components were simple 
versions of the planned flight hardware.  Results on the performance of the tracker, 
calorimeter, and anti-coincidence charged particle veto are presented.}

\baselineskip=20pt

\indent
\vskip 3cm \newpage
\section{Introduction}

GLAST\cite{ref-GLAST} is the next high-energy gamma-ray mission, scheduled to be 
launched by NASA in 2005.  This mission will continue and expand the exploration 
of the upper end
of the celestial electromagnetic spectrum uncovered by the highly 
successful EGRET\cite{ref-EGRET} experiment, which had full sensitivity up to 
$\approx 10$ GeV.  The design of the GLAST instrument is based 
on that of EGRET, which is a gamma-ray pair-conversion telescope, with the primary 
innovation being the use of modern particle 
detector 
technologies.  GLAST will cover the energy range 20 
MeV-300 GeV, with capabilities up to 1 TeV.  It will have 
more than a factor of 30 
times the sensitivity of EGRET in the overlapping energy region (20 MeV - 10 GeV). 
With unattenuated sensitivity to higher than 300 GeV, GLAST 
will cover one of the most poorly measured regions of the electromagnetic spectrum. 
These new capabilities will open important new windows on a wide variety of science 
topics, including some of the most energetic phenomena in Nature: gamma-ray bursts, 
active galactic nuclei and supermassive black holes, the diffuse high-energy gamma-ray
extra-galactic background, pulsars, and the 
origins of cosmic rays.  GLAST will also make possible
searches for galactic particle dark matter annihilations and other  
particle physics phenomena not currently accessible with terrestrial accelerators.  
In addition, GLAST will provide an important overlap in energy coverage with 
ground-based air shower detectors, with complementary capabilities.  GLAST has 
been developed by an interdisciplinary collaboration of high-energy particle 
physicists and high-energy astrophysicists.

The pair-conversion measurement 
principle allows a relatively precise determination of the direction of incident 
photons and provides a powerful signature for the rejection of charged particle 
cosmic ray backgrounds, which have a flux as great as $10^4$ times that of cosmic
gamma-ray  
signals. The instrument consists of three subsystems: a plastic scintillation 
anti-coincidence detector (ACD) to veto incident charged particles, a precision 
converter-tracker to record gamma conversions and to track the resulting
$e^+e^-$ pairs, and 
a calorimeter to measure the energy in the electromagnetic shower. Particles 
incident through the instrument's aperture first encounter the ACD, followed by the 
converter-tracker, and finally the calorimeter.  The technologies selected for the
subsystems are in common use in many high-energy particle physics detectors.  In 
GLAST, the scintillation light from the ACD tiles is collected and transported
by wavelength-shifting fibers to miniature 
photomultipliers. The tracking pair converter section (tracker) has layers of thin 
sheets of lead, which convert the gammas, and the co-ordinates of resulting charged  
tracks are measured in adjacent silicon strip detectors.  The calorimeter is a 
10 radiation length stack of segmented, hodoscopically-arranged CsI crystals, read 
out by photodiodes.  While the basic principles of these components are 
well-understood, adapting them for use in a satellite-based instrument presents 
challenges particularly in the areas of power and mass.  
The tracker, calorimeter
and associated data acquisition system are modular: the baseline instrument comprises
a 5x5 array of 32x32 cm towers.  In addition to simplifying the construction of
the flight instrument, the modularity also allows detailed testing and 
characterization of all critical aspects of detector performance in the   
full, flight-size configuration early in the development program. 

A detailed 
Monte Carlo simulation of GLAST was used to quantify our understanding of these
technology choices and to optimize the design.  To verify 
the results obtained by the computer analysis simple versions of all
three subsystems were constructed and tested together in an electron and tagged 
photon beam in End Station A at SLAC.  The goals of these tests for each of the 
subsystems included the following: \\ \\ \underline{ACD} \begin{enumerate}
\item{Check the efficiency for detecting minimum ionizing particles (MIPs) using fiber 
readout of scintillating tiles.} 
\item{Investigate the backsplash from showers in the calorimeter, which 
causes false vetoes, as a function of energy and angle (this self-veto was the primary 
limitation of the sensitivity of EGRET at high energy).} 
\end{enumerate} 
\underline{Tracker} 
\begin{enumerate}  
\item{Demonstrate the merits of a silicon strip detector (SSD) pair conversion 
telescope.}
\item{Validate the computer modeling and optimization studies with
respect to converter thickness, detector spacing and SSD pitch.} 
\item{Validate the prototype, low power front end electronics used to read out the 
SSDs.}
\end{enumerate}
\underline{CsI Calorimeter} 
\begin{enumerate}
\item{Demonstrate the hodoscopic light sharing concept for co-ordinate measurement in
transversely mounted CsI logs, and validate the shower imaging performance.}
\item{Measure the energy resolution.}
\item{Study leakage 
corrections using longitudinal shower profile fitting at high energies.}
\end{enumerate}

    For each of these tests, the presence of the other subsystems proved valuable:
for tracker studies (particularly at low energies) the calorimeter provided the 
measurement of the photon energy; for calorimeter studies the tracker provided 
a precision telescope to locate the entry point and direction of the beam particle; 
and for all tests the ACD system was used to discard contaminated events 
({\em e.g.,} accompanying low-energy particles coming down the beam pipe).  We report 
here the 
results of these studies.  In section 2, the experimental setup
of the beamline and the detectors is described.  The performance of the 
individual detectors is given in section 3, followed by a compendium of results from 
the studies for each subsystem in section 4 and a summary in section 5.

\section{Experimental setup}
\subsection{Beamline and Trigger}

\begin{figure}[p]
\centerline{\epsfxsize=7.5in \epsfbox{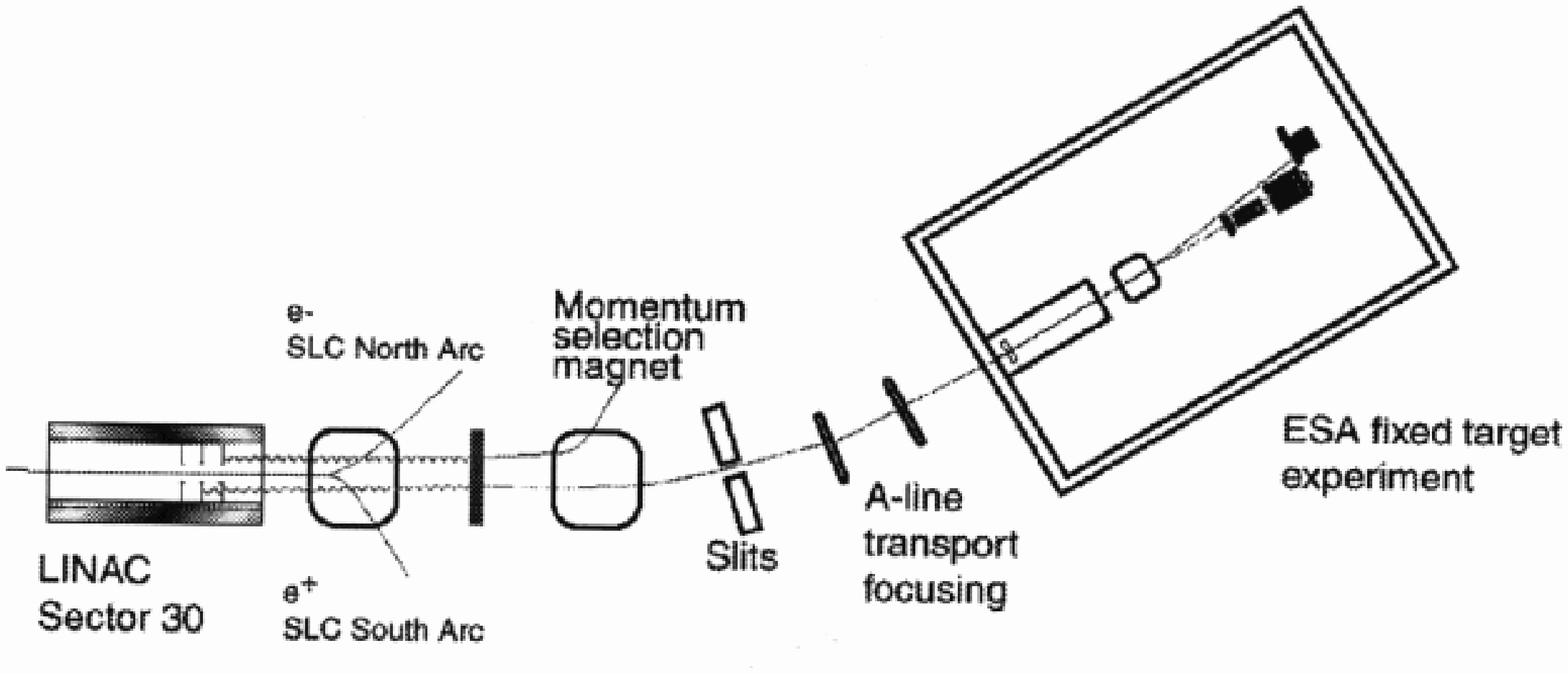}}
\caption{Schematic layout of the SLAC ESA beamline.} 
\label{fig-scrape} 
\end{figure}

The experiment was performed in End Station A (ESA) at the Stanford Linear 
Accelerator Center (SLAC).  A technique was recently 
developed\cite{ref-SLAC_parasitic} to produce relatively low-intensity secondary 
electron and positron beams parasitically from the main LINAC beamline, which 
delivers beams with energy up to 50 GeV at a 120 Hz repetition rate.  A schematic is 
shown in figure \ref{fig-scrape}. A small fraction of the LINAC electron beam is 
scraped by collimators, producing bremsstrahlung photons that continue downstream, 
past bending magnets, producing secondary electrons and positrons when they hit a 0.7
$X_0$ target. Electrons within an adjustable range of momentum (typically 1-2\%) are 
transported to ESA. Beamline parameters were adjusted to allow an average of one 
electron per machine pulse into ESA.

\begin{figure}[p]
\centerline{\epsfxsize=7.5in \epsfbox{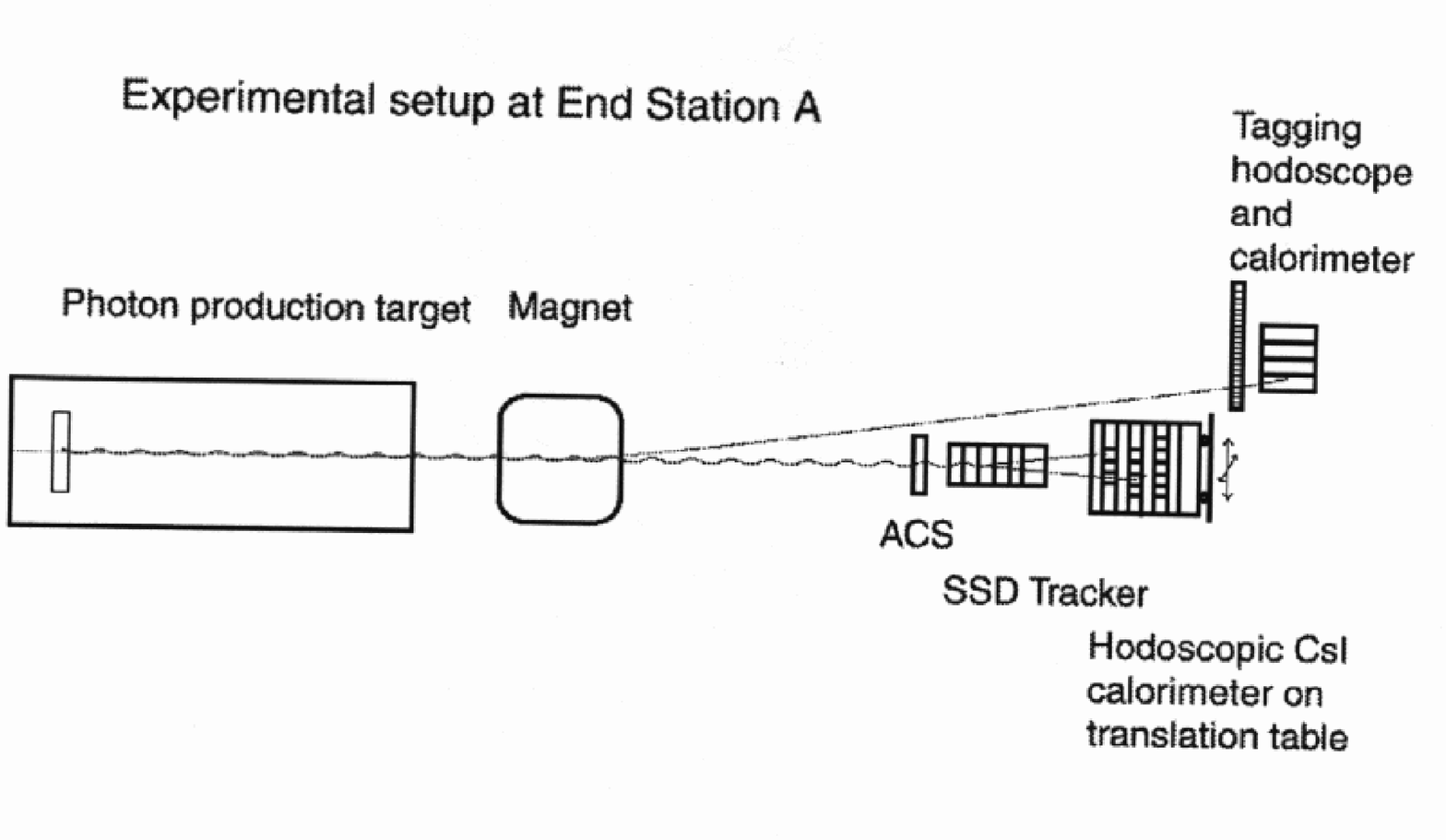} }
\caption{Schematic of the tagged photon beam.} 
\label{fig-tagged_photon} 
\end{figure}

In addition to the electron beam, a tagged photon beam was also generated, as 
shown in figure \ref{fig-tagged_photon}.  A movable target with 2.5\%, 5\% and 10\% 
$X_0$ copper foils produced bremsstrahlung photons from the ESA electron beam (a 25 
GeV ESA electron beam was used for most of the photon runs).  A large sweeping magnet
($B0$) deflected the electron beam toward an 88-channel two-dimensional scintillator 
hodoscope, followed by a set of four lead-glass block calorimeters.

The data acquisition system\cite{ref-Perry} collected data from every machine pulse.
More than 400 data runs were taken during a four week period, resulting in 
$2.1\times 10^8$ triggers and over 200 GB of data.

\begin{figure}[p]
\centerline{\epsfxsize=7.5in \epsfbox{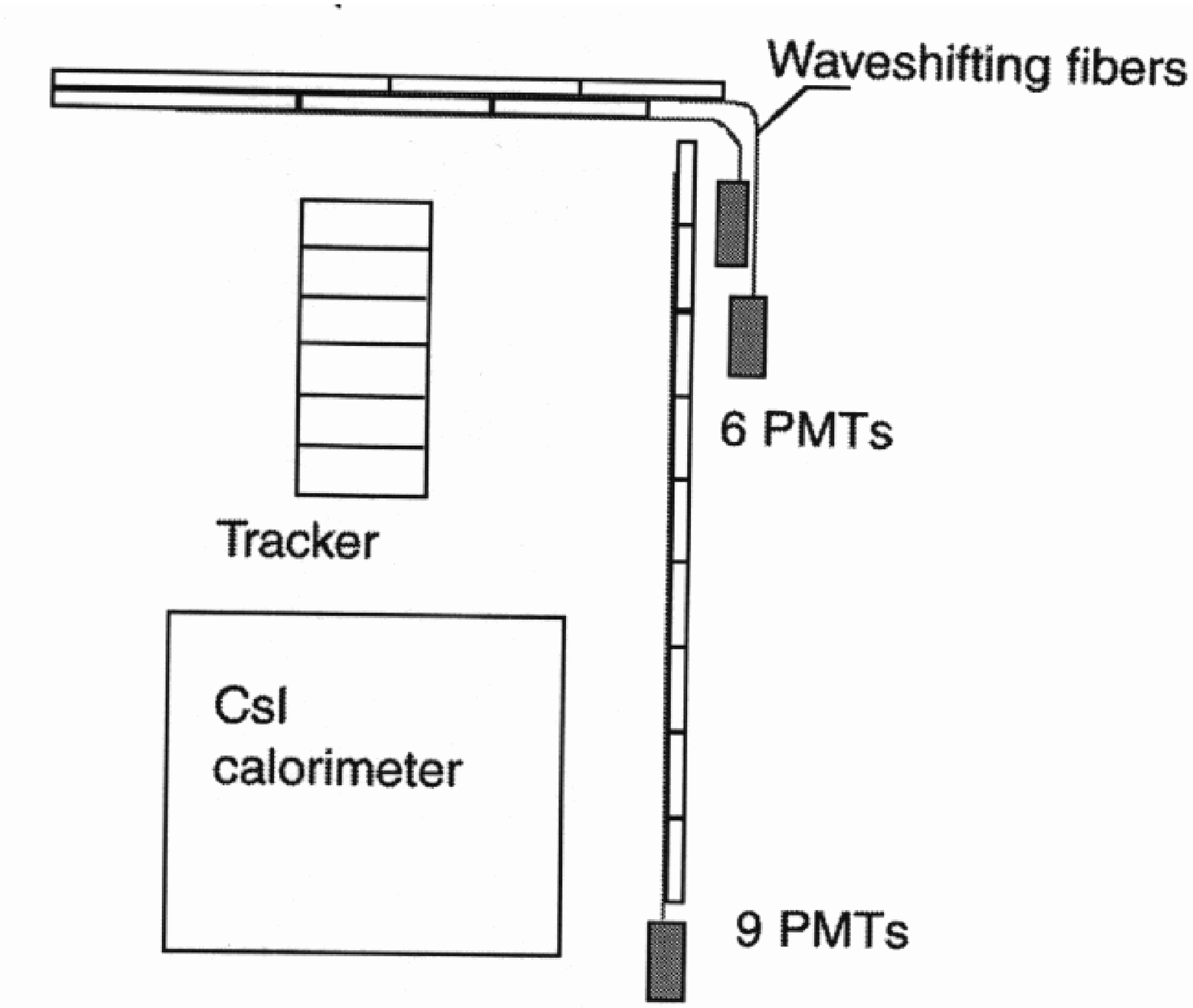} }
\vskip 5cm
\caption{Schematic of the GLAST beam test detectors.} 
\label{fig-setup} 
\end{figure}

The GLAST experimental setup is shown schematically in figure 
\ref{fig-setup}.  Each of the subsystems is described in the following sections.

\subsection{ACD}

Although an anticoincidence system is essential to distinguish the cosmic gamma-ray  
flux from the much larger charged particle cosmic ray flux seen by a gamma-ray 
telescope in orbit, a monolithic scintillator detector such as used by 
SAS-2\cite{ref-SAS2}, COS-B\cite{ref-COSB}, and EGRET is neither practical for an 
instrument the size of GLAST nor desirable. The highest-energy gamma rays (especially
with energies above 10 GeV) produce backsplash: low energy photons originating in the
calorimeter as the products of the electromagnetic shower. Such backsplash photons 
can cause a veto pulse in the ACD through Compton scattering.  The EGRET detector has
a monolithic ACD and suffers a $\approx$50\% loss of detection efficiency at 10 GeV 
due to this effect\cite{ref-Thompson}. This self-veto can be reduced by 
segmenting the GLAST ACD 
into tiles and  vetoing an event only if the pulse appears in the tile through which
the reconstructed event trajectory passes.  Monte Carlo simulations indicate that 
this approach reduces  the self veto rate at 30 GeV by at least an order of magnitude. 

The beam test ACD consisted of two modules, as shown in figure \ref{fig-setup}. 
One module contained 9 scintillating paddles (Bicron BC-408) and was 
placed on the side of the tracker/calorimeter.  The front module consisted of two 
superimposed layers with 3 paddles in each and was placed just upstream of the 
tracker.  
Wave-shifting fibers (BCF-91A, 1 mm diameter), matching the BC-408
scintillator, were embedded in grooves across the 1 cm-thick paddles to collect 
and transfer light to  Hamamatsu R647 photo-multiplier tubes. Each phototube was 
packaged in a soft-iron housing for magnetic field shielding and was
equipped with a variable resistor to adjust the gain. The signal from each 
phototube was pulse-height analyzed by a CAMAC 2249A PHA module. 

\subsection{Tracker}

\begin{figure}[p]
\centerline{\epsfxsize=7.5in \epsfbox{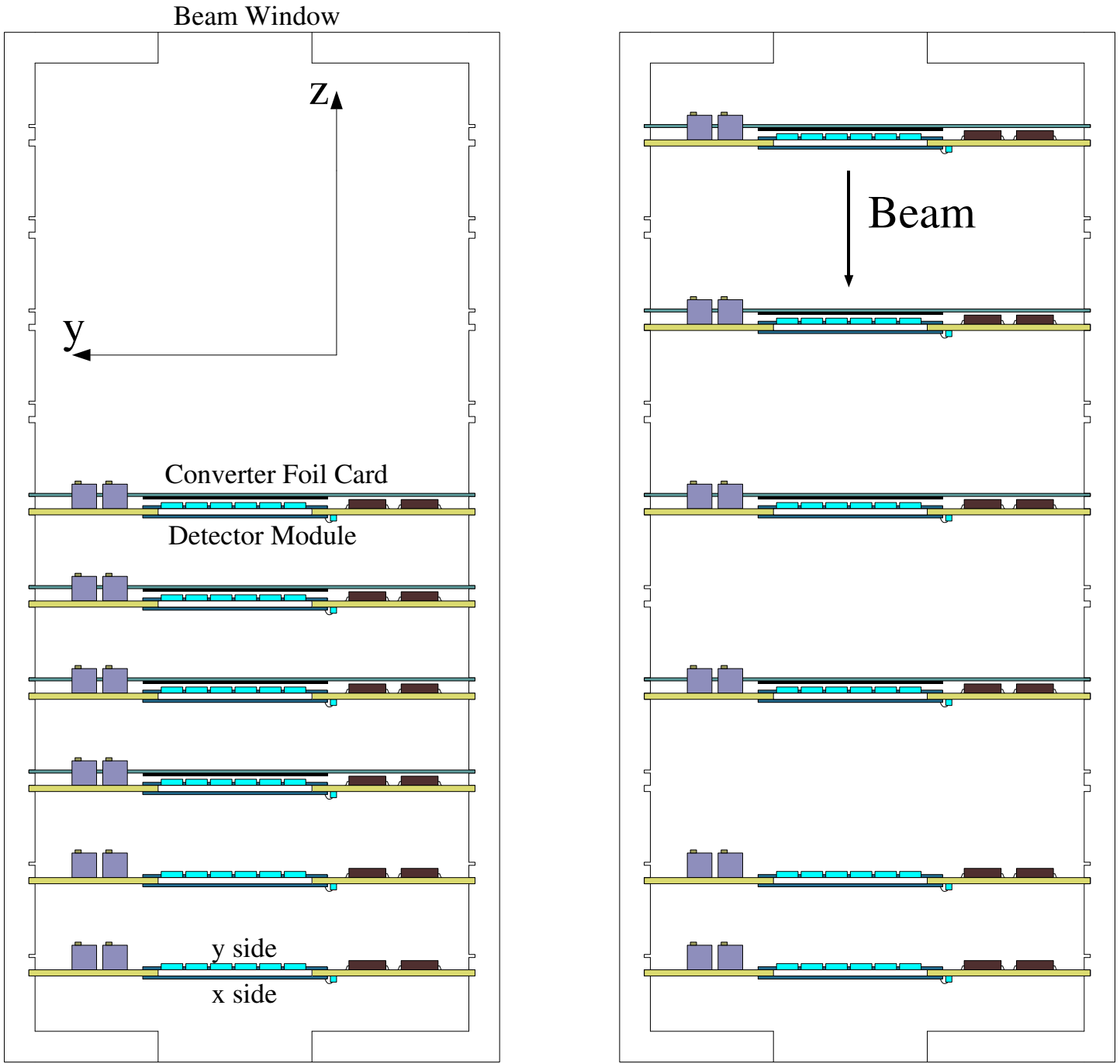} }
\caption{Layout of the beam test tracker.  The ``pancake" configuration is shown on 
the left and the ``stretch" configuration on the right.  
Both configurations are shown with converter foil cards installed in front of the first
four modules.  The $x$ coordinate axis goes into the page.} 
\label{fig-tracker} 
\end{figure}

The silicon-strip tracker consisted of six modules, each with two detector layers, one
oriented to measure the $x$ coordinate and the other oriented to measure $y$.
The detectors were single-sided AC-coupled silicon strip detectors manufactured by
Hamamatsu Photonics.
They were 6~cm by 6~cm in size and 500~$\mu$m thick with an $n$-type substrate and
$p$-type strip implants.
The strips were 57~$\mu$m in width and 236~$\mu$m in pitch, with a total capacitance
of about 1.2~pF per centimeter of strip length.
The strip implants were biased at about 10~V via punchthrough structures, while the
back side was biased at 140~V for full depletion, 
except during special runs in which the bias voltage
was varied in order to study the efficiency as a function of depletion depth.
The detectors were mounted on the two sides of a printed-circuit card, along with the
readout electronics and cable connectors.
To minimize scattering of the beam, each card was cut out under the detector active area, 
and windows were cut out of the acrylic housing.
The entire assembly was wrapped in aluminized mylar for shielding from light
and electromagnetic interference.
Figure \ref{fig-tracker} shows the two general configurations used in the beam test.
The ``pancake" configuration had a 3~cm spacing between modules, similar to the baseline
GLAST design, while in the ``stretch" configuration that spacing was doubled, except for
the space between the last two modules.
The configuration was easily changed by sliding the modules in and out of grooves machined
in the housing.
The lead converter foils were mounted on separate cards that could also be slid in and
out of grooves located directly in front of each detector plane.
In figure \ref{fig-tracker} the converter foils are shown installed in the first four
modules. The gap between the lead and the first detector was about 2~mm, while the 
gap between the two detector sides within a module was 1.5~mm.

Each readout channel
was connected to a single 6~cm long strip, except for the $y$ side of the first 
module encountered by the beam which had five detectors connected in series to 
make 30~cm long strips. Only that module was used for studies of the noise 
performance, since only it had input capacitance 
close to that of the GLAST baseline design.

Consecutive strips were instrumented in each detector with six 32-channel CMOS chips 
that were custom designed to match the detector pitch and satisfy the GLAST power and
noise requirements.  Due to limitations on the number of available readout chips, 
only 192 of the total 256 strips on each detector were instrumented with readout 
electronics. Each channel consisted of a charge-sensitive preamplifier, a shaping 
amplifier with approximately 1~$\mu$s peaking time, a comparator, a programmable 
digital mask, and a latch. In addition, the six chips in each readout section 
provided a 192-wide logical OR of the comparator outputs (after the mask) to provide 
the self-triggering capability required for GLAST.
In the beam test, however, the system was triggered by the beam timing.
About 1~$\mu$s after the beam passed through the apparatus the latches were triggered,
after which the 192 bits were shifted serially out of each readout section.
In addition, the start-time and length of the logical-OR signals 
were digitized by TDC's to study the self-triggering capability offline.

The custom readout electronics operated with a power consumption of 140~$\mu$W per
channel and an rms equivalent noise charge of 1400~electrons (0.22~fC) for the 30-cm 
long strips.
Except for runs in which it was varied to study efficiency, the threshold was generally
set at about 1.5~fC, compared with the more than 6~fC of charge
deposited by a single minimum ionizing particle at normal incidence.
The typical rms variation of the threshold across a 32-channel chip was
under 0.12~fC.
The tracker readout electronics are described in more detail in 
reference \cite{ref-RPJohnson}. 

\subsection{CsI calorimeters} 

The calorimeter comprised eight layers of six 
CsI(Tl) crystals read out by PIN photodiodes.  Each layer was rotated
$90^\circ$ with respect to its neighboring layers, forming an x-y hodoscopic array.  
The crystal blocks were $3 \times 3 \times 19$~cm in size and individually wrapped
in Tetratek and aluminized mylar.  Hamamatsu S3590 PIN photodiodes, with  
approximately 1~cm$^2$ active area, were mounted on each end to measure the 
scintillation light from an energy deposition in the crystal.  The difference in 
light levels seen at the two ends provided a determination of the position of the 
energy deposition along the crystal block.  

\begin{figure}[p]
\centerline{\epsfxsize=5in \epsfbox{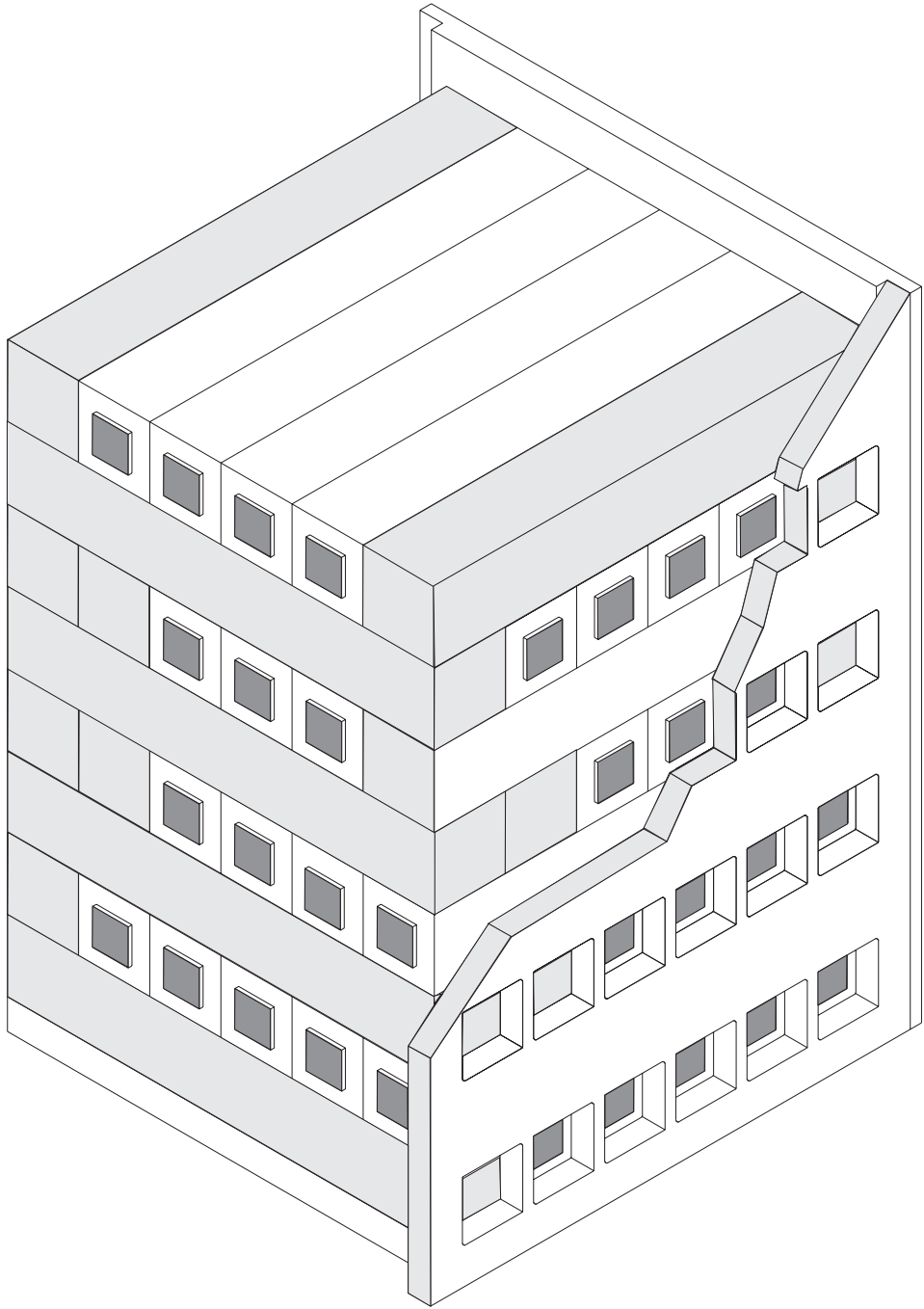} }
\caption{Layout of the beam test calorimeter, as described in the text.} 
\label{fig-calconfig} 
\end{figure}

Although 48 crystals would be required to form the complete 
calorimeter, only 32 CsI(Tl) crystals were available for the test.  Brass
supporting blocks were therefore used to fill the remaining 16 positions to complete 
the hodoscopic array.  Figure~\ref{fig-calconfig} shows the general arrangement of
the calorimeter and the positions of the passive blocks.  In the figure, the
brass blocks are shaded and the CsI blocks are light with PIN photodiodes
indicated on the ends.  The arrangement of the active CsI blocks was designed
to study events normally incident near the front center of the calorimeter.   
Off-axis response could be studied by directing the beam from the front center toward
the lower right corner in the figure where the calorimeter was fully populated with 
active CsI blocks.

The crystal array was mounted in an aluminum frame consisting of four walls
with PIN photodiode access holes and a bottom structural plate.  In 
figure \ref{fig-calconfig},
two of the walls have been removed. The frame was open on the front where the
beam  entered the calorimeter.  The calorimeter was enclosed in a light-tight
aluminum shield and mounted on a precision translation table which permitted both
vertical and horizontal adjustment of the beam position on the front of the
calorimeter.  This translation table was used to study the 
position resolution by mapping the relative light levels
over the entire length of the CsI blocks.  In these tests the tracker remained
fixed and provided accurate beam positions, while the calorimeter was moved
relative to the beam and tracker to map the entire crystal array.

The PIN diodes were biased by 35 volt batteries and attached to eV Products
5092 hybrid preamps.  The preamps were mounted on circuit cards adjacent to the
PIN diodes.  The outputs of the 64 preamps were routed to a CAMAC/NIM data
acquisition system consisting of CAEN shaping amplifiers and Phillips or LeCroy 
12 bit analog to digital converters.  The CAEN shaping amplifiers provided
programmable gain adjustments to optimize the electronics for the specific beam
energies of each test.

\subsection{Online data spying, event display, and offline filtering}

The online system sampled events from
the data stream and made simple data selections 
in real time.  This enabled us to monitor the performance of the
individual detectors and to tune various beam parameters while collecting 
data.
The online monitoring system included a single event display with rudimentary 
track reconstruction and full online histogramming capabilities.  

Offline processing reduced
the volume of data for storage and distribution.
Most of the beam pulses did not result in photon events, due to the thin 
target radiators we used.  To separate real events
from empty pulses we applied very loose selection criteria on the raw data,
requiring either hits in three consecutive x-y tracker planes or at least
6 MeV of energy deposited in the calorimeter.  Event filtering
removed approximately 80\% of the raw data in photon mode and approximately
30\% in electron mode.

\section{Detector performance}

\subsection{ACD}
\begin{figure}[p]
\centerline{\epsfxsize=7.5in \epsfbox{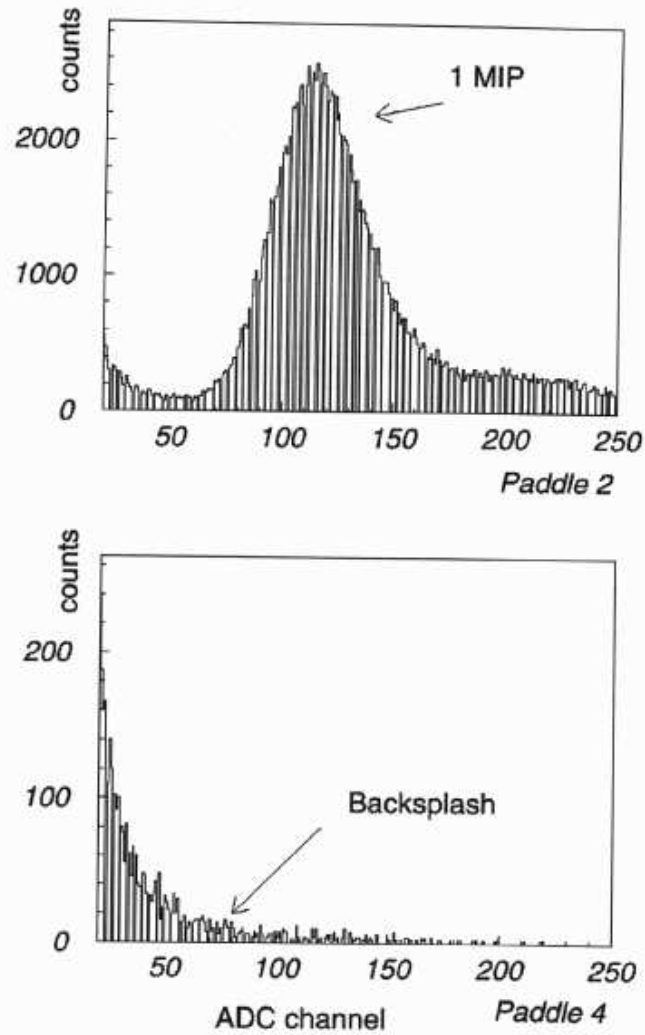} }
\vskip -5cm
\caption{Typical ACD pulse height spectra for (a) a tile that was crossed
by a direct electron beam, and (b) a tile outside the direct beam. The peak
corresponding to one MIP is clearly seen in (a), near channel 100. The backsplash 
spectrum appears in the low channels of histogram (b).} 
\label{fig-acd2} 
\end{figure}

The overall response and efficiency of the ACD were investigated using a 25 GeV 
electron beam.
Typical pulse height histograms are shown in figure \ref{fig-acd2} for 
(a) a tile that was crossed
by a direct electron beam, and (b)  a tile outside  the direct beam. The peak
corresponding to one MIP is clearly seen in (a), near channel 100. The backsplash spectrum
appears in low channels of histogram (b).  

The efficiency was determined using a sample of electron beam events that had 
hits in all 12 tracker planes within 1 cm of the beam axis by counting the fraction 
of these events that had a coincident hit in the relevant ACD tile.
For thresholds below 0.35 MIP, the inefficiencies were always smaller than 
$5\times 10^{-4}$.

\subsection{Tracker}

The efficiency to detect minimum-ionizing particles and the occupancy due to 
random noise were measured.   
The efficiency must be close to 100\%:
to realize the
optimal angular resolution of the device it is crucial not to miss either of the
first two $xy$ pairs of measurements on a track.
The noise occupancy must be low, not only to avoid flooding the data stream but,
more importantly, to avoid saturating the readout system with spurious triggers.
In GLAST, the tracker will be employed in the first-level trigger, which simply 
looks for a coincidence among three consecutive $xy$ pairs of silicon layers. The
rate for this trigger depends very strongly on the occupancy:
with a 1 $\mu$s coincidence window the single-channel noise
occupancy must be less than $10^{-4}$ so that spurious triggers
do not dominate the overall trigger rate.
A major objective of the beam test was to demonstrate that such a low occupancy can
be achieved with the prototype electronics without degrading the detection efficiency.

\subsubsection{Tracker noise occupancy}

\begin{figure}[p]
\centerline{\epsfxsize=7.5in \epsfbox{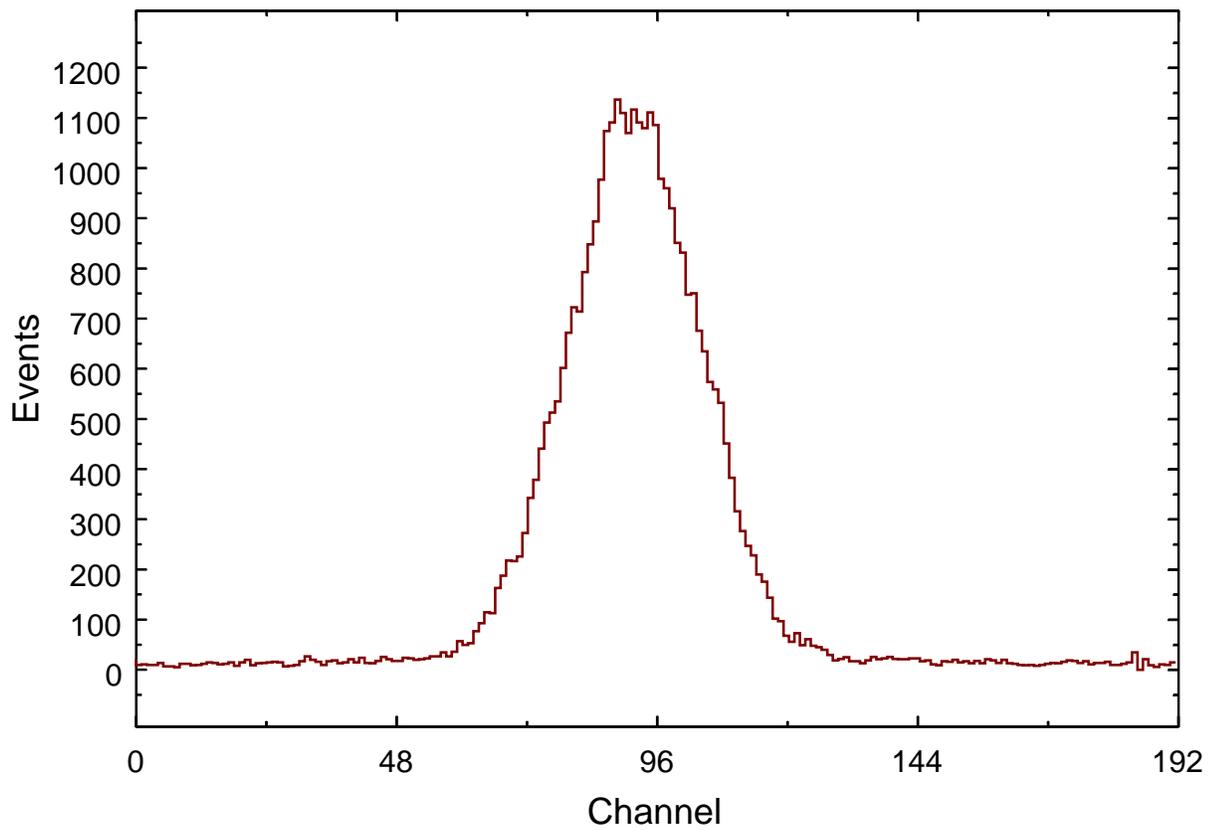} }
\caption{The beam profile as measured in the $y$ coordinate by tracker 
module~4.  The strip pitch is 236 $\mu$m.} 
\label{fig-beamprofile} 
\end{figure}

Only the first layer of detectors struck by the beam had five detectors ganged in 
series, so it was the only relevant testbed for studies of the noise occupancy. 
(Due to poor 
quality control at the wire bonding vendor, a number of detector strips in the
five-detector module were damaged in random locations.  These were known prior to the
beam test and have been removed from the analysis.) The other 
single-detector modules had low capacitance and therefore almost unobservably low 
noise, with the exception of very few damaged strips. 
The efficiency, however, is not 
expected to depend significantly on the capacitance, so it could be studied with the 
single detector modules as well as the first five-detector layer.
Figure~\ref{fig-beamprofile} shows the vertical beam profile. 
It is well contained within the 4.5~cm instrumented region of the detector.
 
\begin{figure}[p]
\centerline{\epsfxsize=7.5in \epsfbox{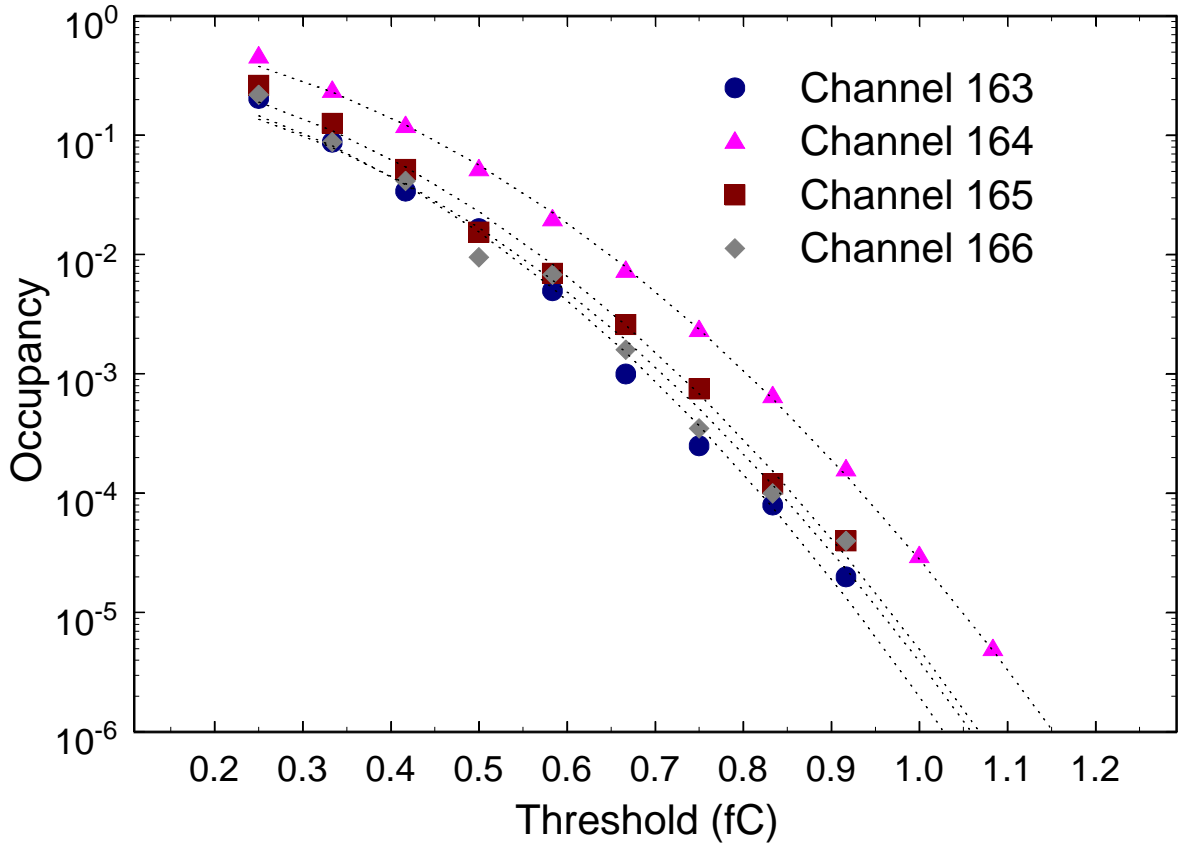} }
\caption{The noise occupancy versus threshold in four channels near the center
of the beam profile in the five-detector module.
The threshold is given in terms of equivalent noise charge at the preamplifier
input.  One fC represents a voltage at the discriminator input of about 115~mV.
The smooth curves are fits to the equation in the text.  From the fits, the rms noise
values of channels 163 through 166 are 1277, 1386, 1315, and 1322 electrons 
respectively.} 
\label{fig-occupancy} 
\end{figure}

In the case that random hits are due to electronic noise, 
the dependence of the threshold-crossing rate, or noise rate, on the threshold level 
$V_t$ is well approximated by~\cite{ref-Rice} 
\begin{equation}
\label{eqn_noise}
f_n = f_0 \cdot e^{-V_t^2/2\sigma_n^2}\, ,  
\end{equation}
where $\sigma_n$ is the rms noise level at the discriminator input. 
Figure~\ref{fig-occupancy} shows the occupancy for four
typical channels of the five detector module. 
For these measurements all channels but one were masked off at the output of
the comparator.
The rms noise is extracted by fitting the curves to Eqn.~\ref{eqn_noise}, with the
results plotted as smooth dotted curves in figure~\ref{fig-occupancy}.
The value of $\sigma_n$ in those four fits ranges from 1290 electrons to
1390 electrons (0.21~fC to 0.22~fC)
equivalent noise charge referenced to the preamplifier input.
The channel-to-channel variation in noise occupancy is primarily due to
threshold variations.
The typical rms variation across a 32-channel chip was
0.05~fC, with a few chips showing rms variations as large as 0.14~fC.

The occupancy increased significantly, however, when the outputs of all channels
were enabled.
In that condition the logical-OR of all channels (Fast-OR) 
---which is to be used for triggering---
runs much faster, and its signal was observed to feed back to the amplifiers, causing
a shift in the effective threshold.
Steps have been taken to solve this problem by improving the grounding and 
and power-supply isolation and decoupling
of the circuit board onto which the chips are mounted; by changing the
CMOS Fast-OR outputs to low-voltage differential signals; and by decreasing
the digital power supply from 5\,V to 3\,V.
A prototype chip and
circuit board fabricated with these new features does not exhibit this feedback
problem---the occupancy no longer depends on the number of enabled channels.

\subsubsection{Tracker efficiency}

\begin{figure}[p]
\centerline{\epsfxsize=7.5in \epsfbox{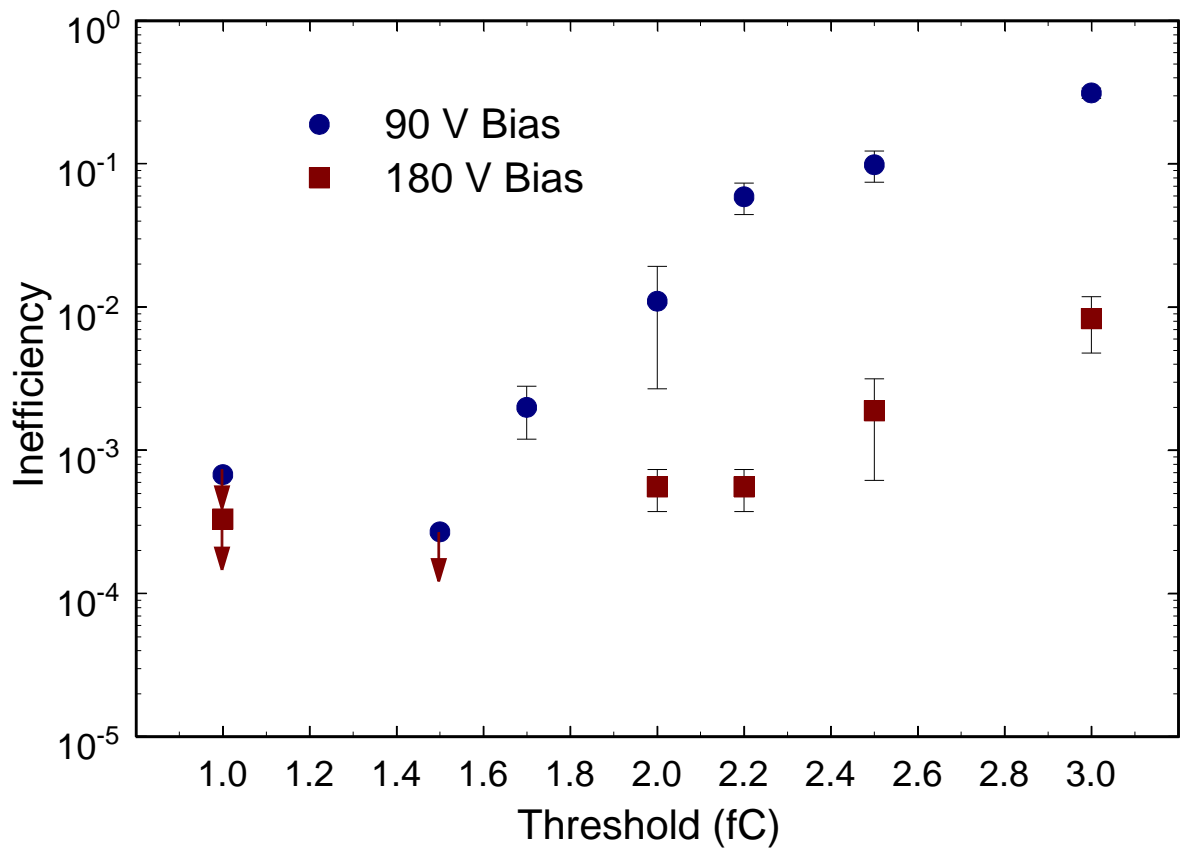} }
\caption{The measured tracker single-plane inefficiency versus threshold setting for two
different bias voltages.  
The measurements were made using events with single 25~GeV electrons.} 
\label{fig-eff} 
\end{figure}

The efficiency was measured for the five-detector module and a single-detector module. 
The remaining four modules were used as anchor planes to 
reconstruct the track. A 25~GeV electron beam was used.
Single particle events were selected by requiring that the
calorimeter signal was consistent with a single-electron shower
and that only one track was reconstructed in the tracker. 
For the detectors under test, a hit was counted if it 
was found within 4 strips of the position 
predicted from the track. 
The bias voltage was varied to change the 
depletion thickness and, therefore, the amount of ionization deposited.  
At about 180 V the 500~$\mu$m detectors were fully depleted. A 
90~V bias voltage yielded a depletion thickness between 360~$\mu$m and 390~$\mu$m, 
close to the envisaged GLAST detector thickness of 400~$\mu$m. 
Figure \ref{fig-eff} shows the inefficiency versus threshold setting for the two 
bias voltages.
The upper limits reflect the limited number of recorded events (about $10^4$). 
No significant difference in efficiency
was observed between the single-detector planes 
and the five-detector plane. 

From figures~\ref{fig-occupancy} and~\ref{fig-eff} it is evident that the tracker can
be operated at essentially 100\% efficiency with an occupancy well below
$10^{-4}$ by setting a threshold in the range of 1-1.5~fC.
The GLAST signal-to-noise and trigger requirements have been met and exceeded, 
while the 140~$\mu$W per channel
consumed by the amplifiers and discriminators satisfies the GLAST
power restrictions. 
More recent tests with prototype chips containing the full GLAST digital readout
capability have demonstrated that, even with the digital activity included,
the per-channel
power can meet the goal of 200~$\mu$W.

\subsubsection{Fast-OR}

The Fast-OR signal was studied in the beam test using multi-hit TDC's.
The distribution of the time of the leading edge is important for understanding
the GLAST trigger timing requirements.
It was measured 
with high-energy electrons for a variety of
detector bias voltages corresponding to depletion depths ranging between 200\,$\mu$m
and 500\,$\mu$m.
For full depletion, the full width of the peak at half maximum was only 50\,ns.
The lower bias voltages resulted in larger time fluctuations, but overall the
data indicated that a trigger coincidence window of 0.5\,$\mu$s could be used for
minimum ionizing particles with essentially 100\% efficiency.

\begin{figure}[p]
\centerline{\epsfxsize=7.5in \epsfbox{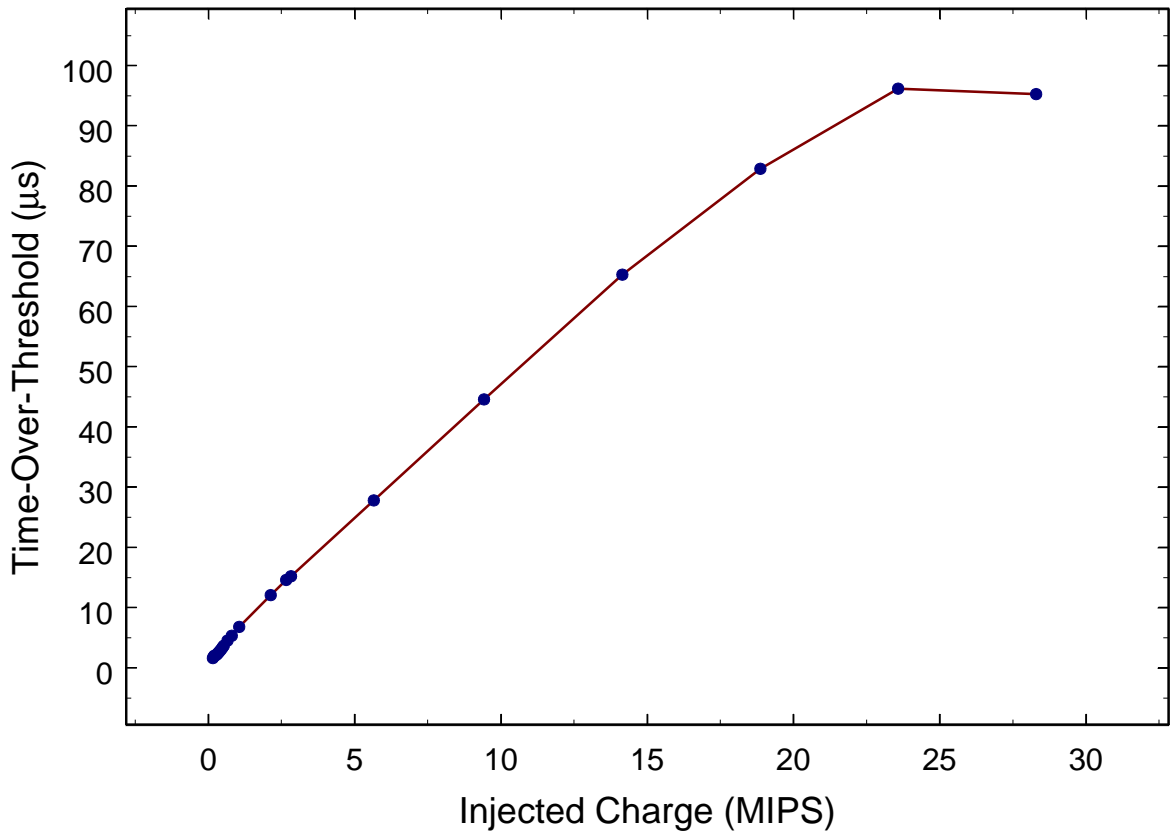} }
\caption{The time-over-threshold of the Fast-OR signal as a function of input charge
injected via the internal calibration capacitors.} 
\label{fig-tot} 
\end{figure}

The GLAST experiment will record the time-over-threshold of the
Fast-OR from each detector layer, along with
the hit pattern.
The time-over-threshold gives a rough measurement of the charge deposited by
the most highly ionizing track that passed through the layer.
That information can be useful for background rejection as well as for possible
cosmic ray studies.
Figure~\ref{fig-tot} shows the measured time-over-threshold versus input signal,
obtained via charge injection, since the beam test did not provide a controlled,
wide range of charge deposition.
The relationship, which would be logarithmic for a true RC/CR filter, is actually
fairly linear in the range 0.5-25~MIPs, where it saturates at 95\,$\mu$s.
This is because, for large amplitudes, the shaping amplifier reset rate is
limited by a constant current source.

\subsection{Calorimeter}

The number of electrons produced in the
photodiodes per MeV deposited energy was measured in several channels of the CsI
calorimeter array.  To calibrate the yield, a known charge was injected into 
each channel and the response was compared with the pulse height distribution
produced by cosmic-ray muons, 
which typically deposit $\simeq$20 MeV in a crystal.  
The yield was typically $\sim$12,000-15,000 electrons per MeV per photodiode 
in the 19-cm CsI bars with a 3 $\mu$s amplifier shaping time.  

\section{Studies} 

\subsection{ACD studies}

\begin{figure}[p]
\centerline{\epsfysize=8.in \epsfbox{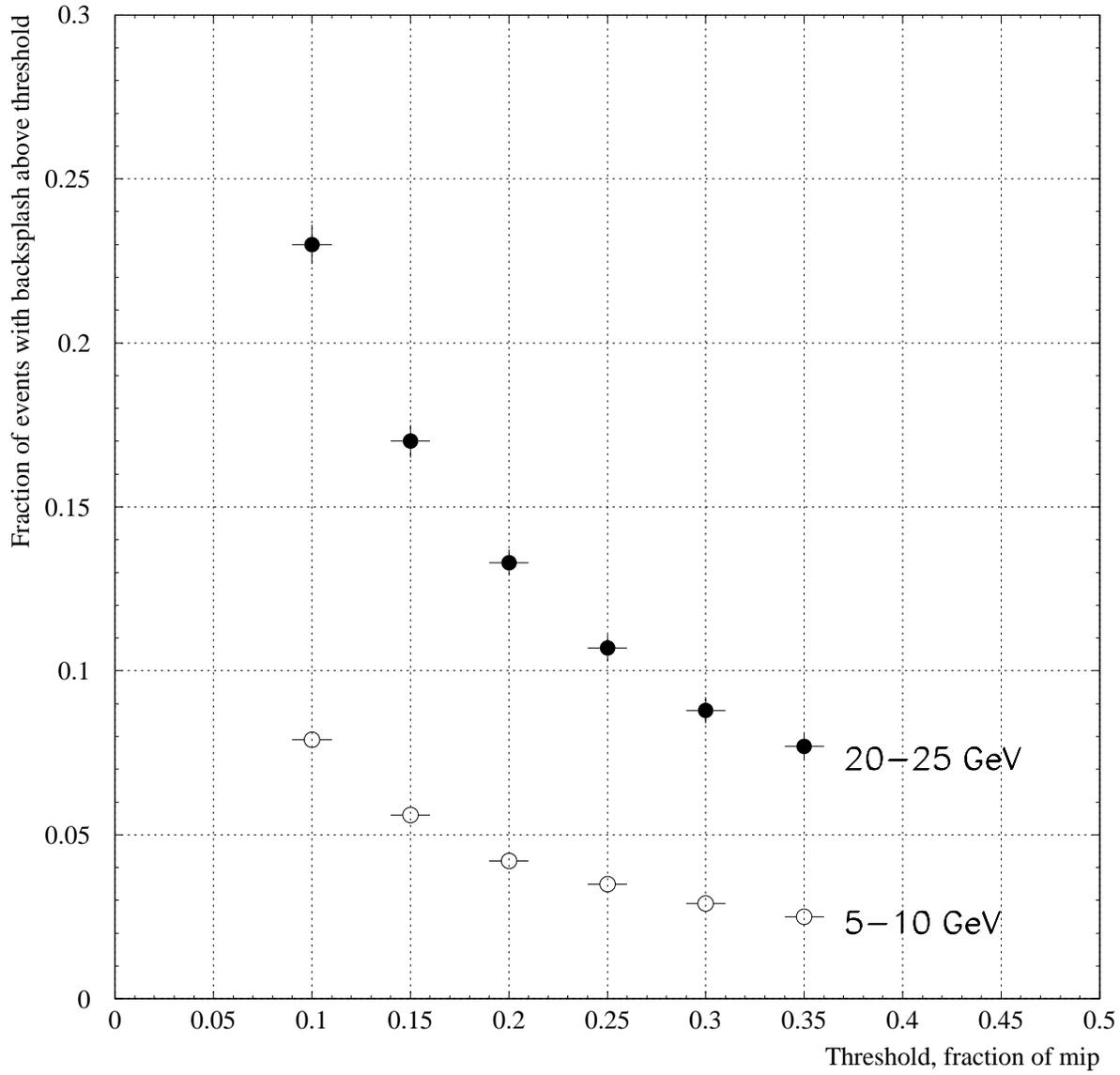} }
\caption{Fraction of events with a backsplash pulse in tile 9,
approximately 90$^{\circ}$ from the incident photon direction, as a function
of threshold for 5-10 and 20-25 GeV photons.} 
\label{fig-backsplashtile9} 
\end{figure}

\begin{figure}[p]
\centerline{\epsfysize=8.in \epsfbox{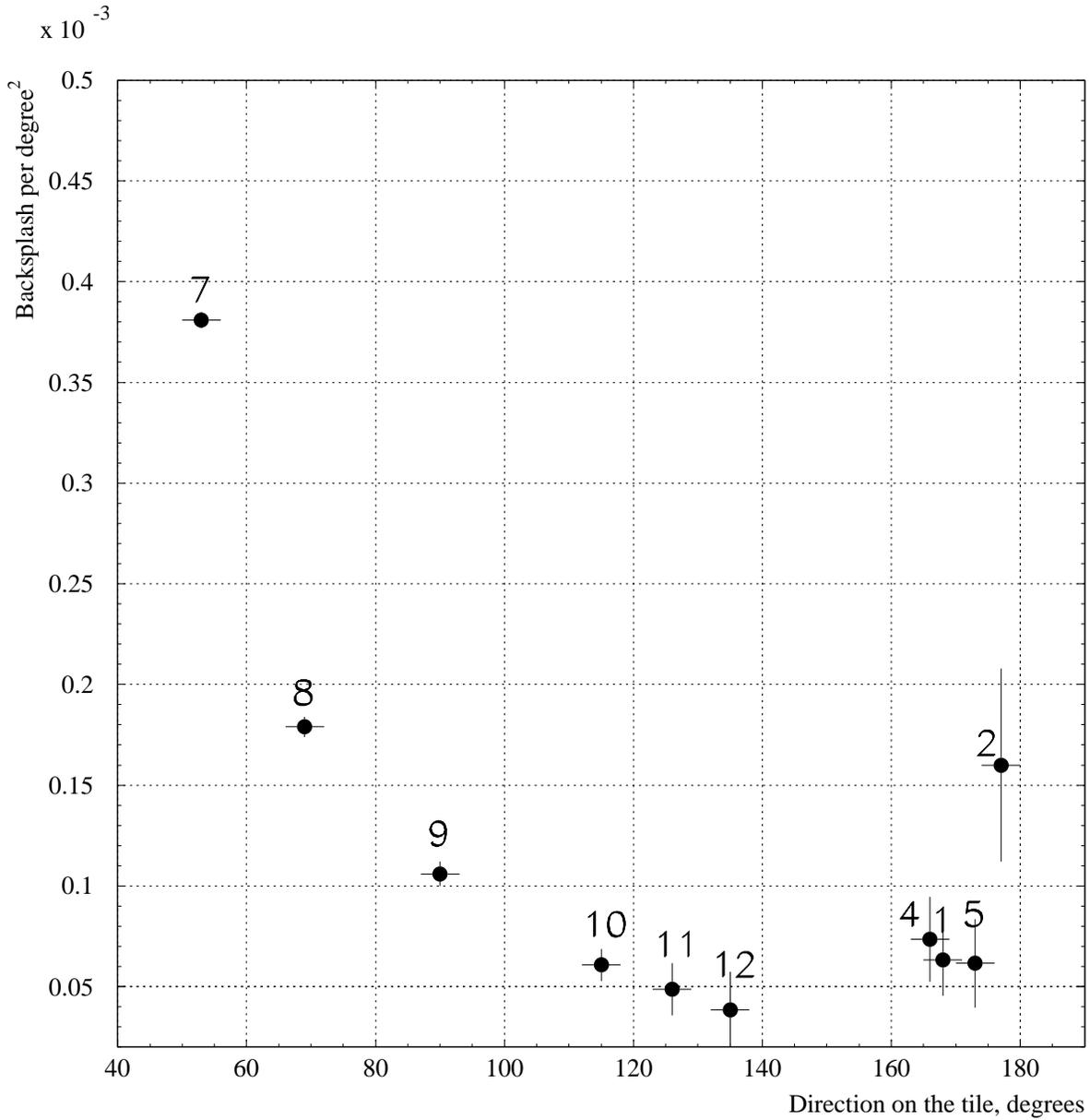} }
\caption{The fraction of events accompanied by an ACD pulse of greater than 
0.2 MIPs as a 
function of angle with respect to the incident photon direction.  The number accompanying
each point indicates the ACD tile number.  The vertical 
axis is normalized by the solid angle each tile presents when viewed from the 
center of the shower in the calorimeter.  Errors are statistical only.} 
\label{fig-backsplashvsangle} 
\end{figure}

The nine scintillator tiles on the side of the tracker/calorimeter and those on the 
top that were not directly illuminated by the beam were used to 
measure the 
backsplash. Figure \ref{fig-backsplashtile9} shows, as a function of threshold,  
the fraction of events that were accompanied by a pulse in tile number 9 which, when
viewed from the center of the shower in the calorimeter,
was approximately 90$^{\circ}$ from the direction of the incident photon.  The 
self-veto effect is a sensitive function of this threshold.  
In figure \ref{fig-backsplashvsangle}, the fraction of events that
were accompanied by an ACD pulse of greater than 0.2 MIP is shown as a 
function of angle with respect to the incident photon direction.  To present
the result in a manner that is insensitive to geometry, the vertical axis is normalized
by the solid angle each tile presents when viewed from the center of the shower in the
calorimeter.  Only the statistical errors are displayed; the systematic errors, 
which may be substantial, are being evaluated.  The increase at 180$^{\circ}$ may 
be due to secondary particles
in the beam accompanying the photon.  Aside from this feature, and the effects
of shower leakage, the backsplash is apparently approximately isotropic.

\subsection{Tracker studies}

\subsubsection{Track reconstruction}

The incident \gammaray ~direction is determined from the electron and 
positron tracks, which are reconstructed from the set of hit strips.
In addition to effects of noise hits, missing hits, and spurious or 
ambiguous tracks, the pointing resolution is ultimately limited by 
hit position measurement error and by energy-dependent multiple 
scattering.
Furthermore, the $x$ and $y$ projections of the instrument are read
out separately so that, given a track in the $x$ projection, the question
of which $y$ track corresponds to it is ambiguous.  Clearly, 
a good method of finding and fitting electron tracks will be
critical for GLAST.

The Kalman filter \cite{bjwft:kalman}
is an optimal linear method for fitting particle
tracks. A practical implementation has been developed by
Fr\"{u}hwirth\cite{bjwft:fruehwirth}.
The problem simplifies
in the limit where either one of the resolution-limiting
effects is negligible: 
if the measurement error were negligible compared to effects of multiple
scattering, as expected at low energies,
the filter would simply ``connect the dots,'' making a track from 
one hit to the next; however, if the measurement error were completely dominant 
and multiple 
scattering effects were negligible ({\em e.g.,} at
high energy), all hits would have information and one would 
essentially fit a straight line to the hits.
The Kalman filter effectively balances these limits.

The basic algorithm we have
adopted is based on the Fr\"{u}hwirth implementation.  At each
plane the Kalman filter predicts, based on the information from the prior 
planes, the most likely location of
the hit for a projected track.  Usually, the hit nearest to that
predicted location is then assumed to belong to the track.  This simple
approach is complicated by opportunities for tracks to leave the
tracker or to share a hit with another track. 
For each event, the algorithm looks for electron tracks in the two 
instrument projection planes ($xz$ and $yz$) independently.  The
fitted tracks are used to calculate the incident photon direction, as 
described in the following sections.

\subsubsection{Simulations}
Simulations of the beam test instrument were made using a 
version of \glastsim\ \cite{ref-Suson} specially modified to represent the beam
test instrument.  \glastsim\ is the code used to simulate the
response of the entire \glast\ instrument via the detailed interactions
of particles with the various instrument and detector components
\cite{ref-GLAST}. 
The Monte Carlo code was modified for the beam test application to include
the \el\ beam, the Cu conversion foil, and the magnet used for analyzing
the tagged photon beam, as well as the 
beam test instrument.  Simulated data were
analyzed in the same way as the beam test data. 

\subsubsection{Cuts on the Data}
Each event used in the analysis was required to pass several cuts.
First, the Pb glass blocks used for tagging must have indicated that there was
only one electron in the bunch.  This lowered the probability of having
multiple \gammarays\ produced at the bremsstrahlung target.  Second, the
Anti-Coincidence tiles through which the \gammaray\ beam passed were each 
required to have less than $1/4$ MIP of energy registered.  This
ensured that the \gammaray\ did not convert inside the ACD tile and that the event 
was relatively clean of accompanying low-energy particles from the beamline.  
Depending on run conditions, this left about 30\% of the data for further analysis. 
Three more cuts were imposed based on the parameters of the reconstructed tracks: 
tracks must have had at least three hits regardless of the energy in the calorimeter,
a reduced $\chi^2 < 5$, and the starting position of the track must have been at 
least $4.7$ mm from the edge of the tracker. This last requirement lowered the 
probability that a track might escape the tracker, which could bias the reconstructed
track directions. These overall track definition cuts further reduced the data by 
about one third.

In an effort to make the \beamtest\ data as directly comparable with 
Monte Carlo simulations as possible, the Monte Carlo data were subjected
to very similar cuts.  The Monte Carlo included an anti-coincidence 
system, and a similar cut was made to reject events which converted in 
the plastic scintillator.  All of the cuts based on track parameters 
were made in the same way for both the Monte Carlo and the 
\beamtest\ data. 

\subsubsection{Reconstructing photon directions}
Since the average pair conversion results in unequal sharing of the \gammaray\ 
energy, and since multiple scattering effects are inversely 
proportional to the energy of the particle producing the track, the 
incident \gammaray\ angles
were calculated using a weighted average of the two track directions, with the 
straighter track receiving 3/4 of the total weight.  The projected instrument 
angular resolution could be measured by examining the distribution of 
reconstructed incident angles. As this distribution had broader tails than a 
Gaussian, the 68\% and 95\% containment radii were used to characterize it. For 
each instrument configuration, these parameters were measured as a function of energy
in ten bands.  The same reconstruction code was used to analyze the Monte Carlo 
simulations, and the distribution widths were compared (figures \ref{fig-sxplots}
and \ref{fig-pxplots}).  The simulated 
distributions show good agreement with the data out to the
95\% containment radius and beyond.

\begin{figure}[p]
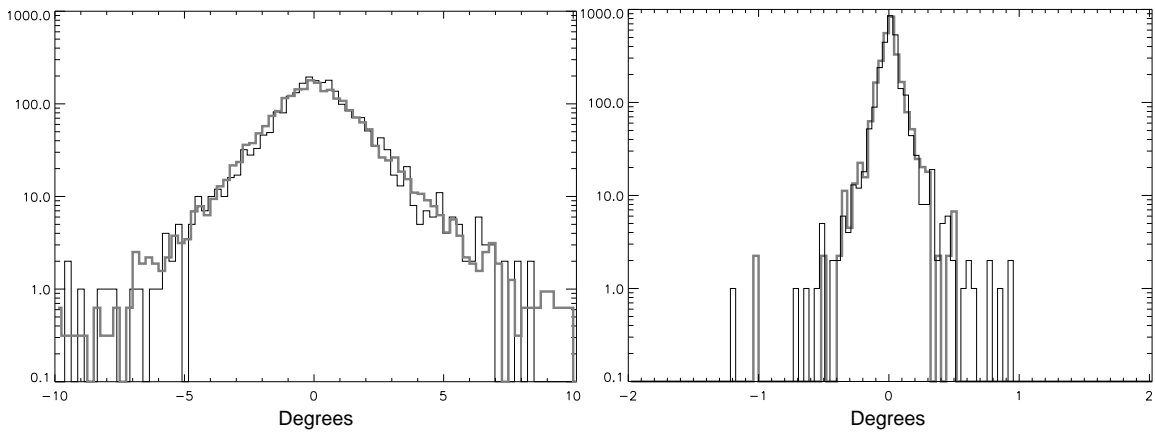

\centerline{\epsfxsize=3.in \epsfbox{atp4xe140MeV.epsi}
\epsfxsize=3.in \epsfbox{ats0xe7.2GeV.epsi}}
\caption{Reconstructed \gammaray\ angle distributions for beam
test and Monte Carlo data for pancake 4\% \radlen\ (left) and stretch with no Pb
radiators (right).  Thin lines are the beam test distributions, 
thick lines are the normalized Monte Carlo distribution.} 
\label{fig-atplots} 
\end{figure}

The containment radii in each projection fall off with increasing
energy somewhat faster than the $1/E$ dependence expected purely from
multiple scattering, for a number of reasons.
The containment radii at low energies are smaller than might be expected 
because of self-collimation: 
the finite width of the detector prevents 
events from being reconstructed with large incident angles.
At higher energies, measurement error becomes a significant 
contributor to the angular resolution.  While these effects
cause deviations from theoretical estimates of the pointing resolution, they are 
well-represented by Monte Carlo simulations (see figure \ref{fig-atplots}).
Details of the angular resolution determination, including specifics
of the track-finding algorithm, methods of dealing with noisy
strips, alignment of the instrument planes, and possible
systematic biases are discussed elsewhere \cite{bjwft:psf}.

\subsection{Calorimeter studies}

\subsubsection{Energy reconstruction}

The principal function of the calorimeter is to measure the energy of incident
$\gamma$-rays.  At the lower end of the sensitive range of GLAST, where
electromagnetic showers are fully contained within the calorimeter, the best
measurement of the incident gamma-ray energy is obtained from the simple sum of
all the signals from the CsI crystals.  At energies above $\sim$1 GeV, an
appreciable fraction of the shower escapes out the back of the calorimeter, and
this fraction increases with $\gamma$-ray energy. At moderate energies ($\sim$
few GeV), fluctuations in the shower development thus create a substantial tail
to lower energy depositions; at higher energies these fluctuations completely
dominate the resolution and the response distribution is again symmetric, but
broader.

\begin{figure}[p]
\centerline{\epsfxsize=5in \epsfbox{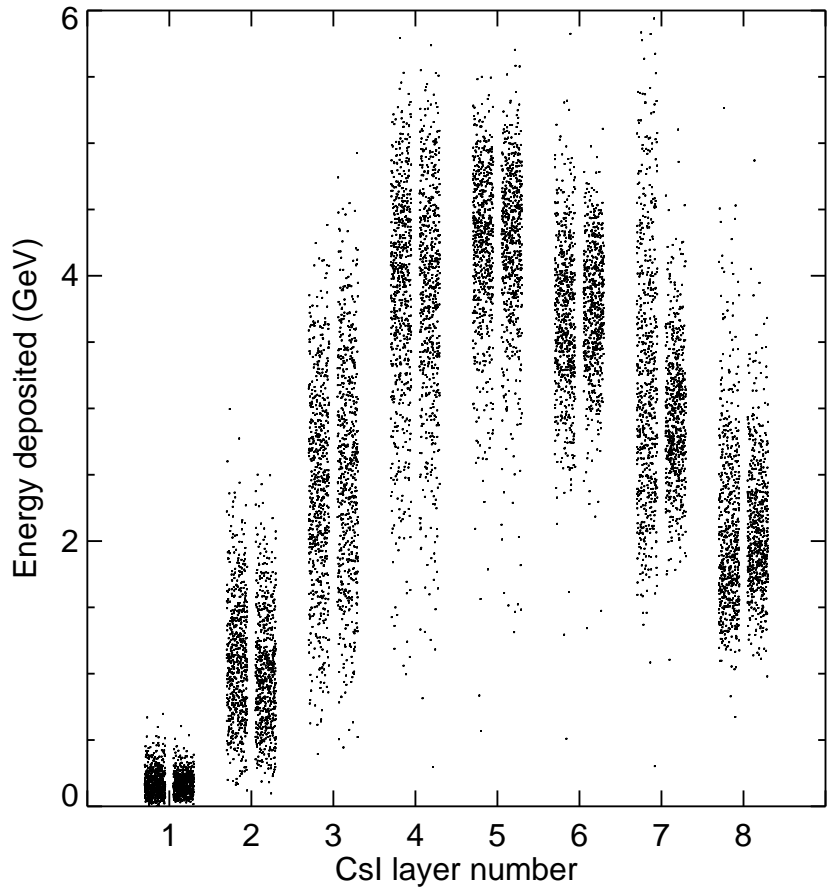} }
\caption
{Energy loss distributions for each of 8 layers in the calorimeter.  There are
two distributions shown per layer: the distribution to the left of the 
layer number is from beam test
events; the distribution to the right is from Monte Carlo simulations.  The
widths of the distributions along the abscissa are arbitrary and serve only to
spread the distribution of energies for display. }
\label{fig-de_distributions}
\end{figure}

Figure \ref{fig-de_distributions} shows the distributions of energy 
deposition for 25 GeV electron showers in each of the 8 layers of the 
beam test calorimeter.  A pair of distributions is shown for each layer: 
the left member of the pair is from the beam test data, with one event producing
one point in each layer; the right member of the pair is the same distribution
from the Monte Carlo simulation.  
The centroid and width of the beam test and Monte Carlo
distributions in each layer are in good agreement quantitatively (with the exception 
of layers 7 and 8, where a configuration error in the ADCs blurred the distributions).
The broad energy distributions seen in the figure are dominated by shower
fluctuations, and the energy depositions are strongly correlated from layer to
layer.  
Using a monoenergetic 160 MeV/nulceon $^{12}C$ beam at the National Superconducting
Cyclotron Laboratory at Michigan State University, the intrinsic energy resolution of 
these CsI crystals with PIN readout was measured to be 0.3\% (rms) at $\approx$ 2 GeV.

Using the longitudinal shower profile provided by the segmentation of 
the CsI calorimeter, one can improve the measurement of the incident electron 
energy by fitting
the profile of the captured energy to an analytical description of the
energy-dependent mean longitudinal profile.  This shower profile is reasonably
well-described by a gamma distribution\cite{ref-Longo} which is a
function only of the location of the shower starting point and the incident
energy $E_0$:
\begin{equation}
   \frac{1}{E_{0}}\frac{dE}{d\xi} =      \\
            \frac{ b ( b \xi) ^{a-1} \mbox{e}^{- b \xi}} { \Gamma (a)} 
   \label{eq-gamma}
\end{equation}
The parameter $\xi$ is the depth into the shower normalized to radiation
lengths,  $\xi = x/X_0$.  The parameter $b$ scales the shower length 
and depends weakly on electron
energy and the $Z$ of the target material; however, a good approximation is
simply to set $ b=0.5$. The parameter $a$ is energy-dependent with the
form $a = 1 + b ( \ln (E_0/E_c ) - 0.5 )$.  $E_c$ is the critical energy where
bremsstrahlung energy loss rate is equal to the ionization loss rate 
($E_c \sim 14$ MeV in CsI).

\begin{figure}[p]
\centerline{\epsfxsize=5in \epsfbox{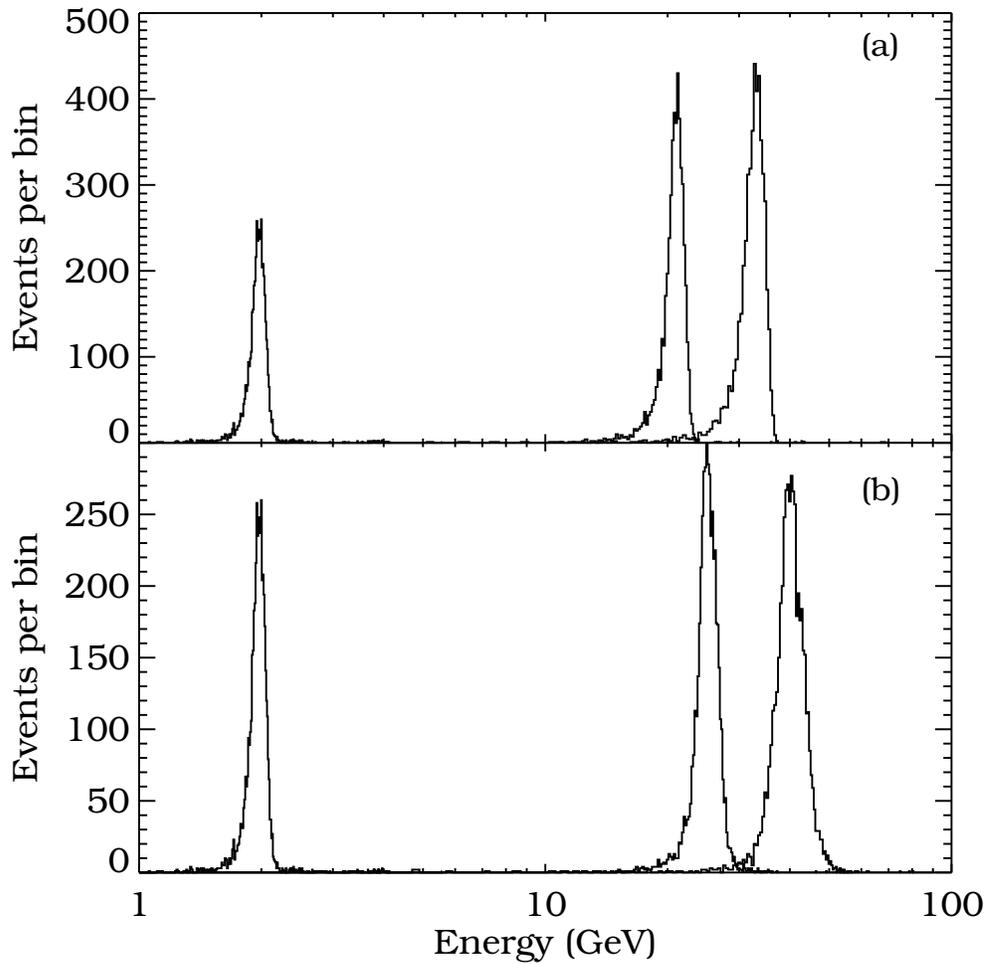} }
\vskip 1cm
\caption{Calorimeter response to 2, 25, and 40 GeV electrons.  Panel (a) displays the
total energy captured in the calorimeter.  Panel (b) shows the results of
longitudinal shower fitting for the 25 and 40 GeV runs as described in the
text.}
\label{fig-calresponse} 
\end{figure}

The  free parameters in the fit were the starting position
of the shower relative to the edge of the first layer of the calorimeter and the
initial electon energy, $E_0$.   In the fitting, the shower profile of Eqn.
\ref{eq-gamma} was integrated over the path length in each of the layers.  The
fitting permitted both early and late starts to the shower.  The results of
the fitting are shown in Figure \ref{fig-calresponse}. Panel (a) of the figure
shows the histograms of  the measured energy loss in the calorimeter for
electron beams of 2, 25, and 40 GeV. The tails to low energy are clearly
evident for the beam energies of 25 and 40 GeV. Figure \ref{fig-calresponse}b 
shows the results of
the fitting as histograms of the fitted  energy for 25 and 40 GeV runs. 
Fitting was not performed for the 2 GeV run and the slight tailing to low
energy is still evident.  The resolutions,  $\sigma_E / E$, as seen
in panel (b) are 4, 7 and 7\% for these three energies.

\subsubsection{Position reconstruction and imaging calorimetry}

The segmentation of the CsI calorimeter allows spatial imaging of the shower 
and accurate reconstruction of the incident photon direction.  Each CsI crystal
provides three spatial coordinates for the energy de posited in it, two
coordinates from the physical location of the bar in the array and one
coordinate along the length of the bar, reconstructed from the difference in
the light level measured in the photodiode at each end ($Left$ and $Right$). To
reconstruct this longitudinal position, we calculate a measure of the light
asymmetry, $A = (Left-Right)/(Left+Right)$, that is independent of the total
energy deposited in the crystal.  We note that if the light attenuation in the
crystal is strictly exponential, the longitudinal position is proportional to
the inverse hyperbolic tangent of the light asymmetry, $x = K \tanh^{-1}A$.

\begin{figure}[p]
\centerline{\epsfxsize=5in \epsfbox{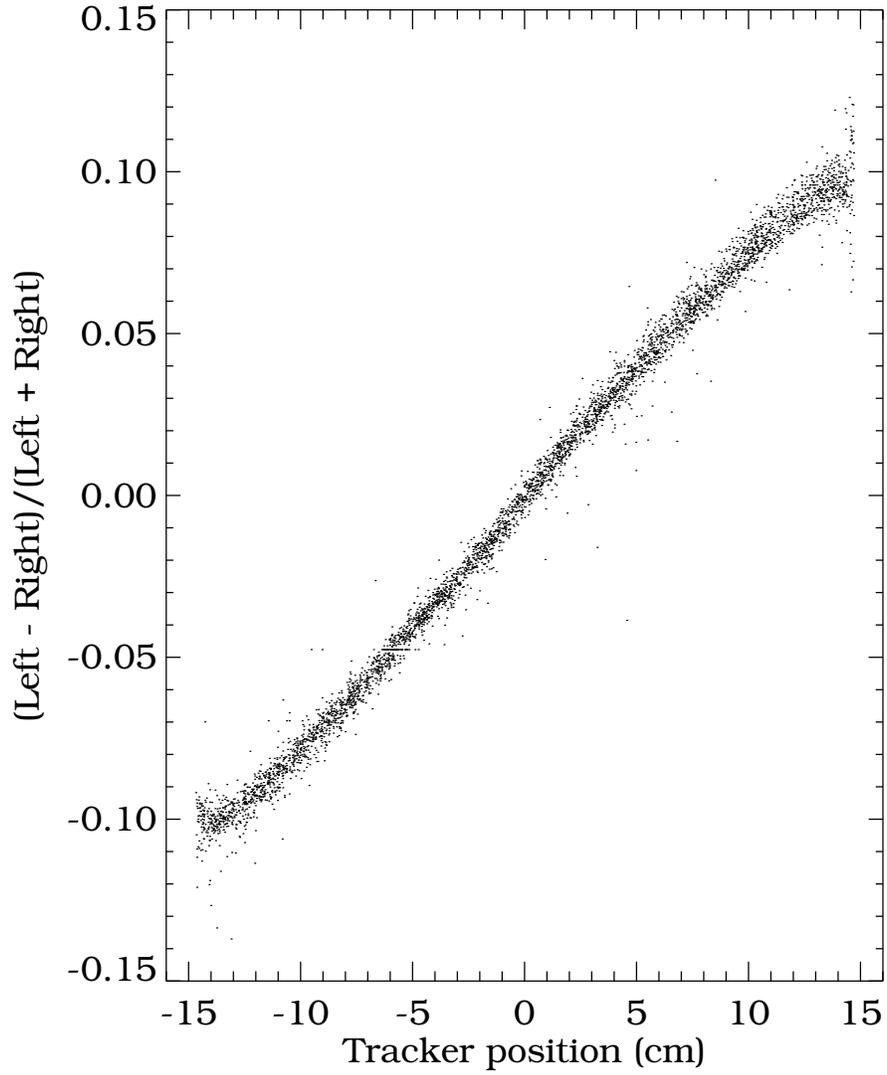} }
\caption{Correlation between the light asymmetry, $A = (Left-Right)/(Left+Right)$,
and the incident 2 GeV electron beam position along the 32-cm CsI bar, as described 
in the text.} 
\label{fig-light_asymmetry} 
\end{figure}

Figure \ref{fig-light_asymmetry} demonstrates that this relationship does
indeed hold in the 32-cm CsI bar, and simple analytic forms can be used to
convert light asymmetry to position.  Positions were determined by the Si
tracker for 2 GeV electrons, which typically deposited $\sim$150 MeV in this
crystal.  The rms error in the position, determined from light asymmetry, is
0.28 cm.

\begin{figure}[p]
\centerline{\epsfxsize=5in \epsfbox{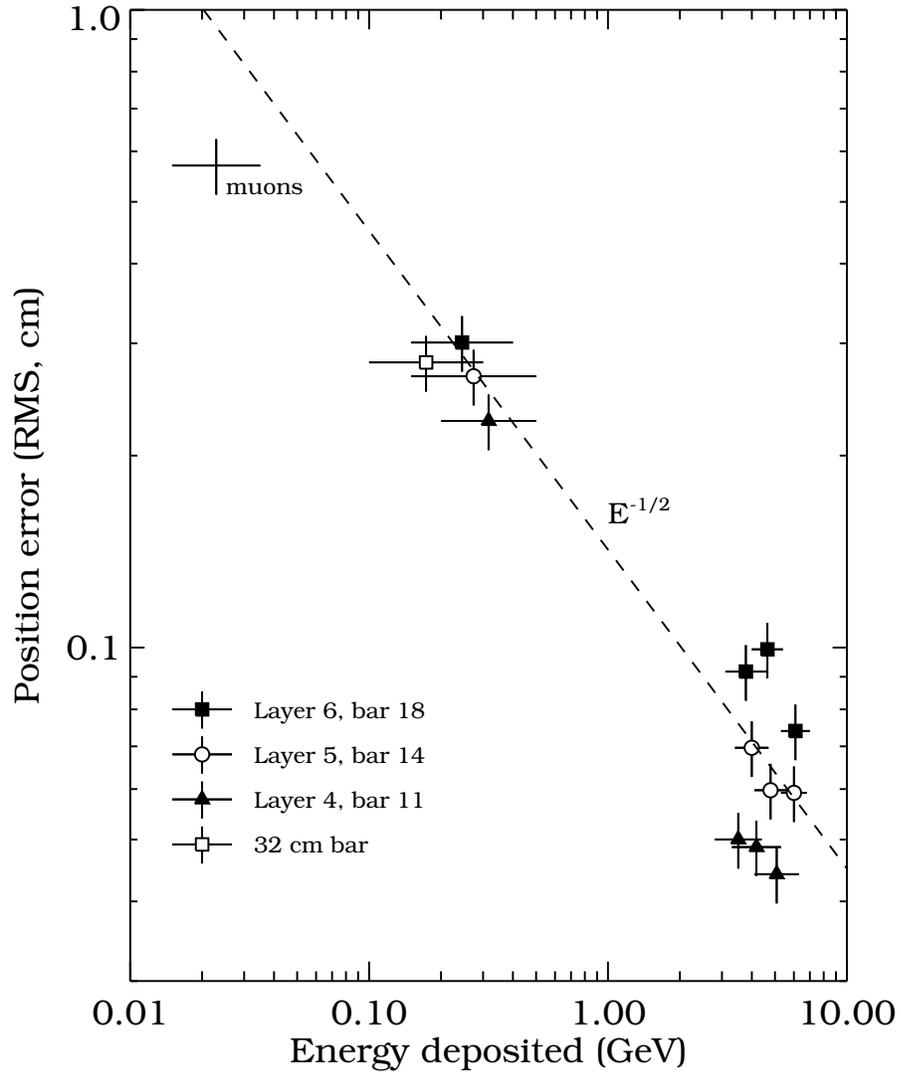} }
\caption{Position resolution, in cm, along the CsI bars for electromagnetic showers and
cosmic ray muons as a function of deposited energy.} 
\label{fig-rms_scaling} 
\end{figure}

The measured rms position error is summarized in the following two figures.
Figure \ref{fig-rms_scaling} shows the position error from three crystals at
increasing depth in the eight-layer CsI array at four beam energies: 2, 25, 30,
and 40 GeV.  The dashed line indicates that the error scales roughly as $1 /
\sqrt E$, indicating that the measurement error is dominated by photon 
statistics.  Also shown is the position error deduced from imaging cosmic-ray
muons in the array, along with that from a 2 GeV electron run in a 32-cm CsI
bar identically instrumented. The muon point falls below those from electron
showers because ionization energy-loss tracks do not have the significant
transverse spread that EM showers have (the Moliere radius for CsI is 3.8 cm).

\begin{figure}[p]
\centerline{\epsfxsize=7.5in \epsfbox{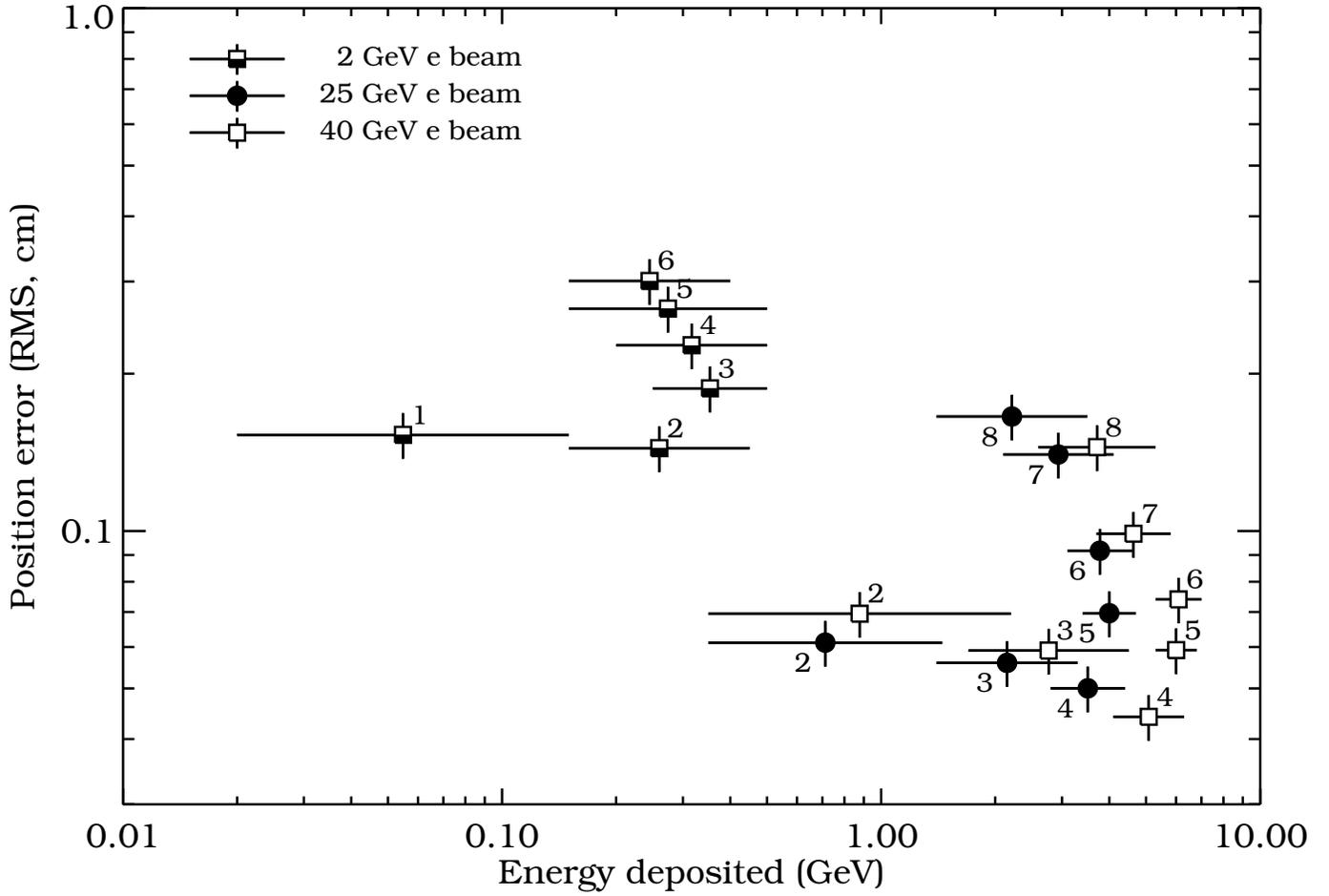} }
\caption{Energy and depth dependence of the position resolution in the calorimeter.
Ordinal numbers indicate the layer in the CsI stack.  The resolution degrades 
significantly after shower maximum.} 
\label{fig-rms_depth} 
\end{figure}

The effect of transverse shower development on position determination can be
seen in figure \ref{fig-rms_depth}.  The rms position error is shown as a
function of energy deposited and depth in the calorimeter (indicated by the
ordinal layer numbers on the data points) for three beam energies.  We see that
the position resolution is best early in the shower, where the radiating
particles are few in number and tightly clustered, and at shower maximum, where
the energy deposited is greatest and statistically easiest to centroid.  The
position resolution degrades past shower maximum, where the shower multiplicity
falls and the energy deposition is spread over a larger area with variations
from shower to shower.

\begin{figure}[p]
\centerline{\epsfxsize=5in \epsfbox{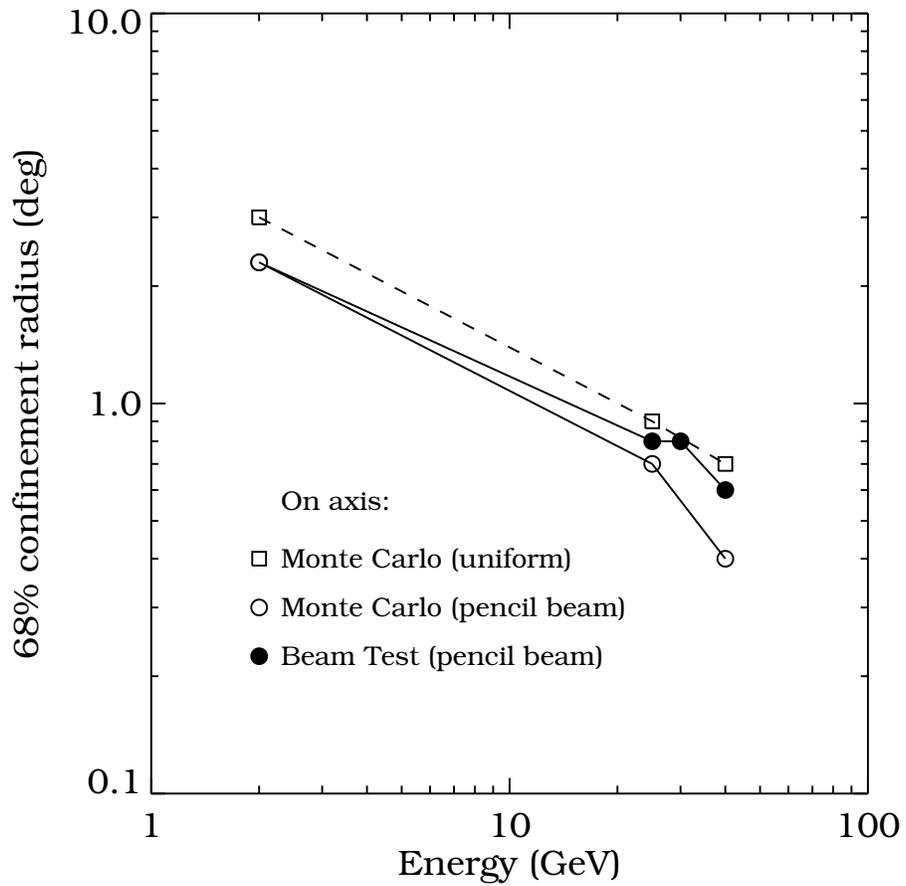} }
\caption{Measured angular resolution using only calorimeter
information, as described in the text, compared with the Monte Carlo simulation.} 
\label{fig-resolution} 
\end{figure}

To test the ability of the hodoscopic calorimeter to image showers, we
reconstructed the arrival direction of the normally-incident beam electrons
from the measured positions of the shower centroids in each layer, without
reference to the tracker information.  The angular resolution, given by the
68\% confinement space angle, is shown by the filled circles in
figure \ref{fig-resolution}.  The open circles indicate the angular
resolution derived from a Monte Carlo simulation of a pencil beam normally
incident and centered on a 3-cm $\times$ 3-cm crossing of crystals.  The
slightly poorer measured resolution is presumably due to systematic errors
in the mapping of light asymmetry to position.  Also indicated, in open
squares, is the angular resolution expected from a uniform illumination
at normal incidence.  Here the angular resolution is degraded because of
transverse sharing of the shower within crystals in a layer of the array.

\section{Summary and Conclusion}
The basic detector elements for GLAST were assembled and tested together for 
the first time in an electron and tagged photon beam at SLAC.  The performance of 
each detector subsystem has been evaluated, and the concept of a silicon strip pair 
conversion telescope has been validated.  The critical tracker performance 
characteristics (efficiency, occupancy, and power) have been investigated in 
detail with flight-size ladders and
meet the requirements necessary for the flight instrument.   Most importantly, 
comparison of the 
results with Monte Carlo simulations confirmed that the same detailed software 
tools that were used to design and optimize the full GLAST instrument accurately 
represent the beam test instrument performance.  A follow-up beam test of a full 
GLAST prototype tower is planned for late 1999.

\section{Acknowledgments}
We thank the SLAC machine group and SLAC directorate for their strong support of this
experiment. 

\clearpage
\section{Figures}

\clearpage

\clearpage

\clearpage

\clearpage

\clearpage

\clearpage

\clearpage

\clearpage

\clearpage

\clearpage

\clearpage

\clearpage

\begin{figure}[p]
\centerline{\epsfxsize=3.in \epsfbox{p0x.epsi}
\epsfxsize=3.in \epsfbox{p4x.epsi}}
\caption{X-Projected angular resolutions for the pancake configuration
with no Pb radiators (left) and 4\% Pb radiators (right).  
Circles indicate the 68\% containment width, and
squares indicate the 95.5\% containment width.
Error bars are 2$\sigma$ statistical errors, and
shaded regions represent the 2$\sigma$ confidence regions of the Monte
Carlo estimates.} 
\label{fig-sxplots} 
\end{figure}

\clearpage

\begin{figure}[p]
\centerline{\epsfxsize=3.in \epsfbox{s0x.epsi}
\epsfxsize=3.in \epsfbox{s4x.epsi}}
\caption{X-Projected angular resolutions for the stretch configuration
with no Pb radiators (left) and 4\% radiators (right).  
Circles indicate the 68\% containment width, and
squares indicate the 95.5\% containment width.
Error bars are 2$\sigma$ statistical errors, and
shaded regions represent the 2$\sigma$ confidence regions of the Monte
Carlo estimates.} 
\label{fig-pxplots} 
\end{figure}

\clearpage

\clearpage

\clearpage

\clearpage

\clearpage

\clearpage

\clearpage

\clearpage

\end{document}